\documentclass[preprint,ascmac,amsmath,amssymb,superscriptaddress]{revtex4-1}
\usepackage{graphicx}
\usepackage{dcolumn}
\usepackage{bm}
\usepackage{color}
\newfont{\bg}{cmr9 scaled\magstep4}
\newcommand{\bigzerol}{\smash{\lower1.0ex\hbox{\bg 0}}}
\newcommand{\bigzerou}{%
\smash{\hbox{\bg 0}}}
\begin{document}
\title{Excitonic optical spectra and energy structures in a one-dimensional Mott insulator demonstrated by 
applying a many-body Wannier functions method to a charge model}
\author{T. Yamaguchi}
\affiliation{Institute of Materials Structure Science,
High Energy Accelerator Research Organization (KEK), 1-1 Oho, Tsukuba 305-0801, Japan}
\author{K. Iwano}
\affiliation{Graduate University for Advanced Studies, Institute of Materials Structure Science,
High Energy Accelerator Research Organization (KEK), 1-1 Oho, Tsukuba 305-0801, Japan}
\author{T. Miyamoto}
\affiliation{Department of Advanced Materials Science, The University of Tokyo, 5-1-5 Kashiwa-no-ha, Chiba 277-8561, Japan}
\author{N. Takamura}
\affiliation{Department of Advanced Materials Science, The University of Tokyo, 5-1-5 Kashiwa-no-ha, Chiba 277-8561, Japan}
\author{N. Kida}
\affiliation{Department of Advanced Materials Science, The University of Tokyo, 5-1-5 Kashiwa-no-ha, Chiba 277-8561, Japan}
\author{Y. Takahashi}
\affiliation{Department of Chemistry, Faculty of Science, and Graduate School of Chemical Sciences and Engineering, 
Hokkaido University, Sapporo 060-0810, Japan}
\author{T. Hasegawa}
\affiliation{Department of Applied Physics, The University of Tokyo, 7-3-1 Hongo, Bunkyo-ku, Tokyo 113-8656, Japan}
\author{H. Okamoto}
\affiliation{Department of Advanced Materials Science, The University of Tokyo, 5-1-5 Kashiwa-no-ha, Chiba 277-8561, Japan}
\affiliation{AIST-UTokyo Advanced Operando-Measurement Technology Open Innovation Laboratory,
National Institute of Advanced Industrial Science and Technology, Chiba 277-8568, Japan}
\date{\today}
%
\begin{abstract}
We have applied a many-body Wannier functions method to theoretically calculate an excitonic optical conductivity 
spectrum and energy structure in a one-dimensional (1D) Mott insulator at absolute zero temperature with large system size. 
Focusing on full charge fluctuations associated with pairs of a holon and doublon, we employ a charge model, which is interpreted 
as a good effective model to investigate photoexcitations of a 1D extended Hubbard model at half-filling in the spin-charge 
separation picture. 
As a result, the theoretical spectra with appropriate broadenings qualitatively reproduce the recent experimental data of 
ET-F$_{2}$TCNQ at 294 K with and without a modulated electric field. Regarding the excitonic energy structure, we have found 
that the excitons, especially for even-parity, are weakly bound by many-body effects. This is also consistent with the fitting 
parameters reported in the recent experiment. 
Thus, our theoretical method presented in this paper 
is practically useful to understand physical roles of charge fluctuations in many-body excited states of a 1D Mott insulator. 
\end{abstract}
\maketitle
\section{Introduction}
\label{sect1}
\par
Recent progress on femtosecond (fs) pulse laser has provided the platform of tunable light-induced excitations 
for various materials in ultrafast timescales and such photoinduced phenomena have been much 
attracting attention toward 
the applications in optical devices and memories\cite{PIPT1,PIPT2,PIPT3}. So far, as a typical example, 
a photoinduced Mott-insulator-to-metal transition has been already observed 
in one-dimensional (1D) Mott insulators of [Ni(chxn)$_{2}$Br]Br$_{2}$ (chxn=cyclohexanediamine)\cite{intr3}, 
ET-F$_{2}$TCNQ (ET=bis(ethylenedithio)tetrathiafulvalene, TCNQ=tetracyanoquinodimethane)\cite{intr4,intr5}, 
[Pd(en)$_{2}$Br](C$_{5}$-Y)$_{2}$H$_{2}$O (en=ethylenediamine, C$_{5}$-Y=dialkylsulfosuccinate)\cite{intr5-2}, and 
Ca$_{2}$CuO$_{3}$\cite{1DMott6} as well as in 
two-dimensional (2D) Mott insulators of Nd$_{2}$CuO$_{4}$, La$_{2}$CuO$_{4}$\cite{intr6,intr7}, and 
$\kappa$-(ET)$_{2}$Cu[N(CN)$_{2}$]Br\cite{intr7-1,intr7-2}. Such nonequilibrium 
photoinduced metallic states are theoretically thought to be attributed to a carrier doping of photogenerated 
holon-doublon (HD) pairs\cite{LgLv,intr8,intr9,intr10}. 
Very recently, the realization of photoinduced $\eta$-pairing superconducting state from a 1D Mott insulator ground state 
has been theoretically predicted\cite{etaP}. 
In the early stage of above phase transitions, pure electronic effects, which are specified by transfer integrals 
of electrons $T$ in the order of 0.1 eV, are considered to be dominant. Then, the time scales of occurring the transitions are 
estimated as on the order of 10 fs. 
One of the powerful experimental tools to investigate such early dynamics immediately after light irradiation is to observe 
pump-probe or transient absorption spectra by utilizing fs pulse laser. 
To physically understand the spectral features, the nature of all the low-energy electronic excited states, particularly related to 
charge fluctuations, should be precisely revealed in a Mott insulator. 
\par
As a first step to theoretically understand above excited states, we previously proposed a charge model as a novel effective model 
for a 1D extended Hubbard model at half-filling in the spin-charge separation picture with full charge fluctuations related to 
HD pairs\cite{CHRM}. 
So far, at least 1D level, the spin-charge separation picture is experimentally believed\cite{SCS1} and theoretically established 
in the limit of strong Coulomb interaction strengths\cite{SCS2,SCS3,SCS4,SCS5,SCS6,SCS7,Charge1}. The ground state is, 
of course, a Mott insulator\cite{Hub1}. 
The completeness of a charge model was confirmed by comparing the optical conductivity spectra, $\sigma(\omega)$, 
for given photon energy $\omega$, between the charge model and Hubbard model in the realistic parameters within the framework 
of the exact diagonalization method\cite{ED1,ED2}. However, this method never accesses to the 
calculations of sufficiently large system size comparable to experiments due to computational problem in principle. 
To theoretically calculate $\sigma(\omega)$ at such large system size with containing many-body effects, 
we also previously proposed a novel method, called a many-body Wannier functions (MBWFs) method\cite{CHRM}. 
The resultant $\sigma(\omega)$ were in good agreement with spectra at huge size computed by dynamical 
density-matrix renormalization group (DDMRG) method\cite{DDMRG1} and time-dependent DMRG (tDMRG) method\cite{tDMRG}. 
\par
Using the MBWFs method, one can also easily interpret, for instance, the physics of all the photoexcited states corresponding to 
the peaks of $\sigma(\omega)$ at sufficiently large size, where finite size effects are negligible. 
Here, to theoretically analyze such excited states in general, non-perturbative schemes such as exact diagonalization method 
and DMRG method\cite{DMRG,rvDMRG} are only permitted because of strong correlations arising from strong Coulomb interactions 
between electrons. At present, 
several reliable $\sigma(\omega)$ in a 1D extended Hubbard model at half-filling with huge size have already been calculated 
by DDMRG scheme\cite{DDMRGJ,DDMRG1,DDMRG2,DDMRG3,DDMRG4,DDMRG5,DDMRG6}. However, in that scheme, because 
the highly renormalized wavefunctions of both the ground state and photoexcited states are obtained as {\it numerical data}, 
to convert the data into corresponding wavefunctions of the {\it real} system size is far difficult. In contrast, our MBWFs method 
can directly obtain the {\it real} wavefunctions with tunable degrees of many-body effects at arbitrary system size. 
\par
In this paper, as a second step, we construct an effective model of the even-parity low-energy excited states associated with 
a one pair of holon (H) and doublon (D) in a 1D Mott insulator by applying a MBWFs method to a charge model. 
This is because our previous MBWFs approach, to theoretically calculate ordinary $\sigma(\omega)$, 
only constructed an effective model of the odd-parity low-energy one-HD-pair photoexcited states from a 1D Mott insulator 
ground state having even-parity. However, to completely understand all the low-energy excitations of a 1D Mott insulator, 
the even-parity excited states with a single HD pair are also important. Regarding this, the excitonic energy structures of 
including information of both the odd- and even-parity low-energy one-HD-pair excited states can be estimated by 
electroreflectance spectroscopy with teraheltz (THz) electric fields. 
So far, several experiments have reported such excitonic energy structures in Mott insulators deduced from fitting parameters 
of measured third-order nonlinear susceptibility, $\chi^{(3)}(-\omega;0,0,\omega)$\cite{1DMott1,1DMott3,1DMott4,1DMott5}. 
As one of such experiments, we choose the recent experiment of ET-F$_{2}$TCNQ at 294 K\cite{1DMott1} and reproduce 
its excitonic energy structure for evaluating the completeness of our even-parity effective model in this paper. 
In this experiment, $\chi^{(3)}(-\omega;0,0,\omega)$ was obtained from a nonlinear polarization 
$P(\omega)\propto \chi^{(3)}(-\omega;0,0,\omega)E^{2}(\omega\sim 0)E(\omega)$, where 
$E(\omega\sim 0)$ can be regarded as a static electric field created by a THz pulse and $E(\omega)$ is an electric field of 
optical probe pulse, respectively. From the fitting analysis with a four-level model for the spectra, which are proportional to 
$E^{2}(\omega\sim 0){\rm Im}\chi^{(3)}(-\omega;0,0,\omega)$, the excitonic energy structure was determined 
as fitting parameters. 
Here, ET-F$_{2}$TCNQ is well known as one of the quasi-1D conductors and the 1D chain consisting of ET molecules 
is aligned in the direction to $a$-axis. The electronic model is well described by a 1D extended Hubbard model at half-filling 
due to negligibly weak electron-lattice coupling\cite{intr4,ETs5}. The typical value of the electronic transfer integral is 
$T\sim0.1$ eV\cite{ETs6}. 
\par
In addition, above experimental spectrum of $\omega E^{2}(\omega\sim 0){\rm Im}\chi^{(3)}(-\omega;0,0,\omega)$ itself is related 
to $\Delta\sigma(\omega)$, which represents the change of $\sigma(\omega)$ between with and without a modulated electric field. 
Then, directly introducing an external modulated electric field to our effective model within the framework of the MBWFs method, 
we also theoretically calculate $\Delta\sigma(\omega)$ and reproduce the corresponding experimental spectrum in this paper. 
This paper is mainly organized in following two sections. 
In Sect.\ref{sect2}, we briefly introduce a charge model and practical applications of MBWFs method to construct the effective 
models of both one-HD-pair odd- and even-parity excited states including the manner of size extension. 
Using the MBWFs, we also demonstrate the scheme of theoretically calculating $\sigma(\omega)$ with and without 
a modulated electric field for large system size. 
All the theoretical spectra and excitonic energy structures compared with corresponding experimental data of ET-F$_{2}$TCNQ 
at 294 K\cite{1DMott1} are shown and discussed in Sect.\ref{sect3}. 
Throughout this paper, we treat absolute zero temperature and set $\hbar=e=c=a=1$, where $a$ is a lattice constant. 
\section{FORMULATION}
\label{sect2}
\par
We treat a half-filled one-dimensional (1D) chain of even $N$ sites with an equal population of spins 
($N_{\uparrow}=N_{\downarrow}=N/2$). First of all, we introduce a Hamiltonian, $H$, and charge-current operator, $J$, 
based on a 1D extended Hubbard model under the periodic boundary condition (PBC) as follows. 
\begin{equation}
H \equiv H_{0} + H_{V} + H_{\phi}, 
\label{eq1}
\end{equation}
\begin{align}
H_{0} &= -T\sum_{j=1}^{N}\sum_{\sigma=\uparrow,\downarrow}\left[
c_{j+1,\sigma}^{\dagger}c_{j,\sigma} + c_{j,\sigma}^{\dagger}c_{j+1,\sigma}
\right] + U\sum_{j=1}^{N}n_{j,\uparrow}n_{j,\downarrow}, 
\label{eq2} \\ 
H_{V} &= \sum_{\alpha}V_{\alpha}
\sum_{j=1}^{N}\sum_{\sigma,\sigma^{\prime}=\uparrow,\downarrow}n_{j+\alpha,\sigma}n_{j,\sigma^{\prime}}, 
\label{eq3} \\
H_{\phi} &= -\phi_{0}\sum_{j=1}^{N}\sum_{\sigma=\uparrow,\downarrow}\sin\left(
\frac{2\pi}{N}(j-1)
\right)
n_{j,\sigma},  
\label{eq4} \\
J &= - iT\sum_{j=1}^{N}\sum_{\sigma=\uparrow,\downarrow}[
c_{j+1,\sigma}^{\dagger}c_{j,\sigma} - c_{j,\sigma}^{\dagger}c_{j+1,\sigma}]. 
\label{eq5}
\end{align} 
Here, $c_{j,\sigma}^{(\dagger)}$ denotes the annihilation (creation) operator of an electron with spin $\sigma=\uparrow, \downarrow$ 
at the $j$th site, $n_{j,\sigma}\equiv c_{j,\sigma}^{\dagger}c_{j,\sigma}$, and $c_{N+1,\sigma}^{(\dagger)}=c_{1,\sigma}^{(\dagger)}$. 
$H_{0}$ is a conventional Hubbard model with a transfer integral $T$ and on-site Coulomb interaction strength $U$. 
According to previous works on ET-F$_{2}$TCNQ\cite{intr9,1DMott1,ETs1}, 
we set $U/T=10$. $H_{V}$ describes a long-range Coulomb interaction and $V_{\alpha}$ corresponds to 
the $\alpha$th nearest neighbor Coulomb interaction strength. In this paper, we consider two special cases, 
\begin{equation}
V_{1}\equiv V, V_{\alpha\geq 2} = 0
\label{eq6}
\end{equation}
and 
\begin{equation}
V_{\alpha}\equiv V/\alpha \; (\alpha=1,2,3), V_{\alpha\geq 4} = 0. 
\label{eq7}
\end{equation}
The former is just a conventional extended Hubbard model and several candidates of $V$ for ET-F$_{2}$TCNQ 
are reported\cite{ETs1,ETs2,ETs3,ETs4}. 
In addition, the latter is treated in the recent paper trying to explain the experimental spectra of ET-F$_{2}$TCNQ at 294 K 
with exact diagonalization analysis at $V/T=4.286$ and $N=14$\cite{1DMott1}. 
However, each above $V$ intrinsically should be determined by finding the best $V$ of reproducing optical conductivity 
spectra measured at sufficiently low-temperature with negligible electron-phonon scattering. 
\par
$H_{\phi}$ represents a periodic scaler potential carrying the momentum $2\pi/N$ and the minus sign is just attributed to 
an elementary charge of electron ($-e=-1$ in this paper). This term generates a modulated electric field and, 
for $2\pi/N\ll 1$, satisfies with an anti-commutation relation $\{ H_{\phi}, \mathcal{P} \} = 0$, where $\mathcal{P}$ denotes 
a parity-inversion operator. $H_{\phi}$ also breaks the translational symmetry. In the experiment of ET-F$_{2}$TCNQ 
at 294 K\cite{1DMott1}, a modulated electric field is introduced by a terahertz pulse in approximately one period of the order 
of 1 picosecond. 
Generally defining a peak magnitude of such experimental modulated electric field as $E_{\rm amp}$/(kV/cm), 
we associate it with $\phi_{0}$ by calculating the root of square mean of a modulated electric field over a single period. 
Referring an unit cell length in the direction of $a$-axis of ET-F$_{2}$TCNQ, 5.791 \AA\cite{ETs5}, we can calculate 
\begin{equation}
\frac{\phi_{0}}{T} = N\frac{9.21666\times 10^{-6}E_{\rm amp}/({\rm kV/cm})}{T/{\rm eV}}\equiv N\varepsilon
\label{eq8}
\end{equation}
for given $T/{\rm eV}$. Hence we also determine the $T$ value in the unit of eV by an experimental lowest excitation energy with 
odd-parity, which corresponds to the energy at the striking peak of an optical conductivity spectrum. 
\par
To determine above $V$ and $T$, we basically calculate optical conductivity spectra defined as 
\begin{equation}
\sigma(\omega) = -\frac{1}{N\omega}{\rm Im}\left[
\langle g|J\frac{1}{\omega+i\gamma+E_{g}-H}J|g\rangle
\right], 
\label{eq9}
\end{equation}
where $|g\rangle$ describes the ground state of $H$ in Eq.(\ref{eq1}) with its energy $E_{g}$ within the framework of 
the linear response theory. $\omega>0$ represents a single photon energy injected into the system (weak photoexcitations). 
In this paper, we also define the maximum value of $\sigma(\omega)$ as $\sigma_{\rm max}$ when $H_{\phi}$ in Eq.(\ref{eq4}) 
is absent, namely in the case of $\varepsilon=0$ in Eq.(\ref{eq8}), for convenience. At absolute zero temperature, 
although $\gamma$ strictly should be an infinitesimal small positive number, we remain this as a certain spectral broadening 
to reproduce the experimental spectrum at finite temperature. 
\par
\begin{figure}[t]
\begin{center}
\includegraphics[width=10cm,keepaspectratio]{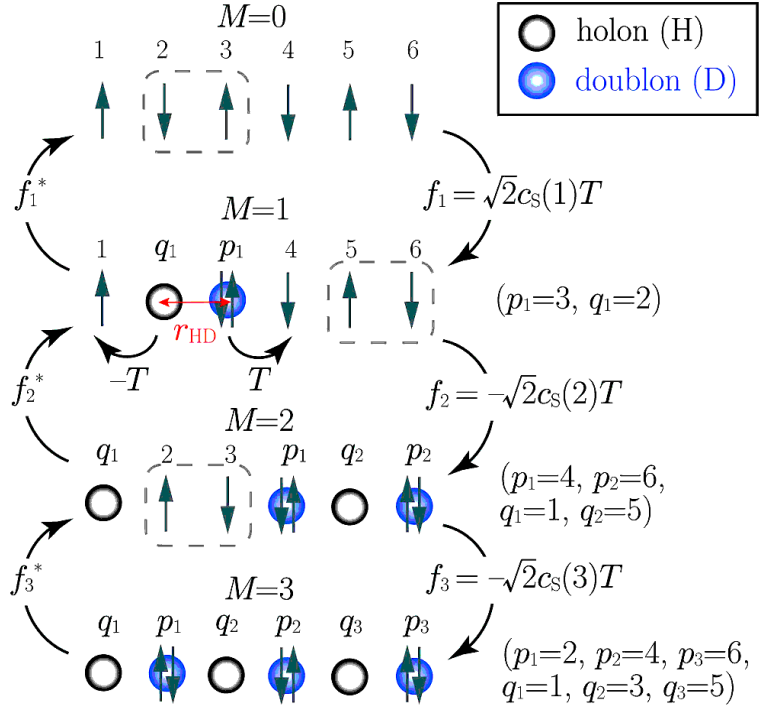}
\caption{Schematics of bare bases $|r_{M}\rangle$ in Eq.(\ref{eq12}) of a charge model with $N=6, M_{\rm max}=N/2=3$. 
A basis $|r_{M}\rangle$ contains $M$ pairs of a holon (H) and doublon (D) with the ground state of a $N-2M$ sites Heisenberg model. 
Up (down) arrows denote up (down) spins. $r_{\rm HD}$ represents a relative distance between a H and D, 
which is defined in the $M=1$ subspace. 
}
\label{fig1}
\end{center}
\end{figure}
\par
\par
Next, we convert $H$ of Eq.(\ref{eq1}) into a charge model, which is an effective model of including full charge fluctuations 
of a 1D extended Hubbard model at half-filling in the spin-charge separation picture as follows. 
\begin{align}
H_{X}^{({\rm C})} &\equiv \sum_{M,M^{\prime}=0}^{M_{\rm max}}P_{M}H_{X}P_{M^{\prime}} \quad (X=0,V,\phi), 
\label{eq10} \\
J^{({\rm C})} &\equiv \sum_{M,M^{\prime}=0}^{M_{\rm max}}P_{M}JP_{M^{\prime}}. 
\label{eq11}
\end{align} 
Here, $P_{M}$ is a projection operator onto the subspace consisting of the basis with $M$ ($1\leq M_{\rm max}\leq N/2$) pairs 
of a holon (H) and  doublon (D), 
\begin{equation}
|r_{M}\rangle\equiv |\{p_{1},p_{2},\cdots,p_{M}\},\{q_{1},q_{2},\cdots,q_{M}\}\rangle, 
\label{eq12}
\end{equation}
where $p_{j}$ ($q_{j}$) represents the site number of the $j$th D (H). Schematics are shown in Fig.\ref{fig1}. 
The rest $N-2M$ sites are singly occupied by electronic spins and are assumed as the ground state of the 1D Heisenberg model. 
Details including derivation of the model was explained in our previous paper\cite{CHRM}. 
Because the Hilbert space is mainly labeled by $M$, we introduce that maximum value, $M_{\rm max}$, 
for convenience in our theoretical calculations. 
Note that $M_{\rm max}=N/2$, of course, corresponds to the full charge fluctuations. 
However, as mentioned in our previous paper, this model has two essential corrections, described 
as the parameters $c_{\rm S}(M)$ and $\theta_{M}$\cite{Charge1,Charge2,Charge3}. 
The former is the correction of the transfer integral corresponding to 
the creation and annihilation of H-D pairs. We set this as $c_{\rm S}(M)=0.82$ in this paper as well as in our previous paper. 
The latter is related to the total momentum of the ground state. In order to set it to zero for all the calculations with even $N$, 
we treat $\theta_{M}=0$ ($\pi$) for odd (even) $M$. Coulomb interaction terms and $H_{\phi}$ in Eq.(\ref{eq4}) have 
no corrections. Consequently, the entire form of our starting model is written as 
\begin{equation}
H^{({\rm C})}\equiv H_{0}^{({\rm C})} +  H_{V}^{({\rm C})} + H_{\phi}^{({\rm C})}. 
\label{eq13}
\end{equation}
\par
\begin{figure}[t]
\begin{center}
\includegraphics[width=7cm,keepaspectratio]{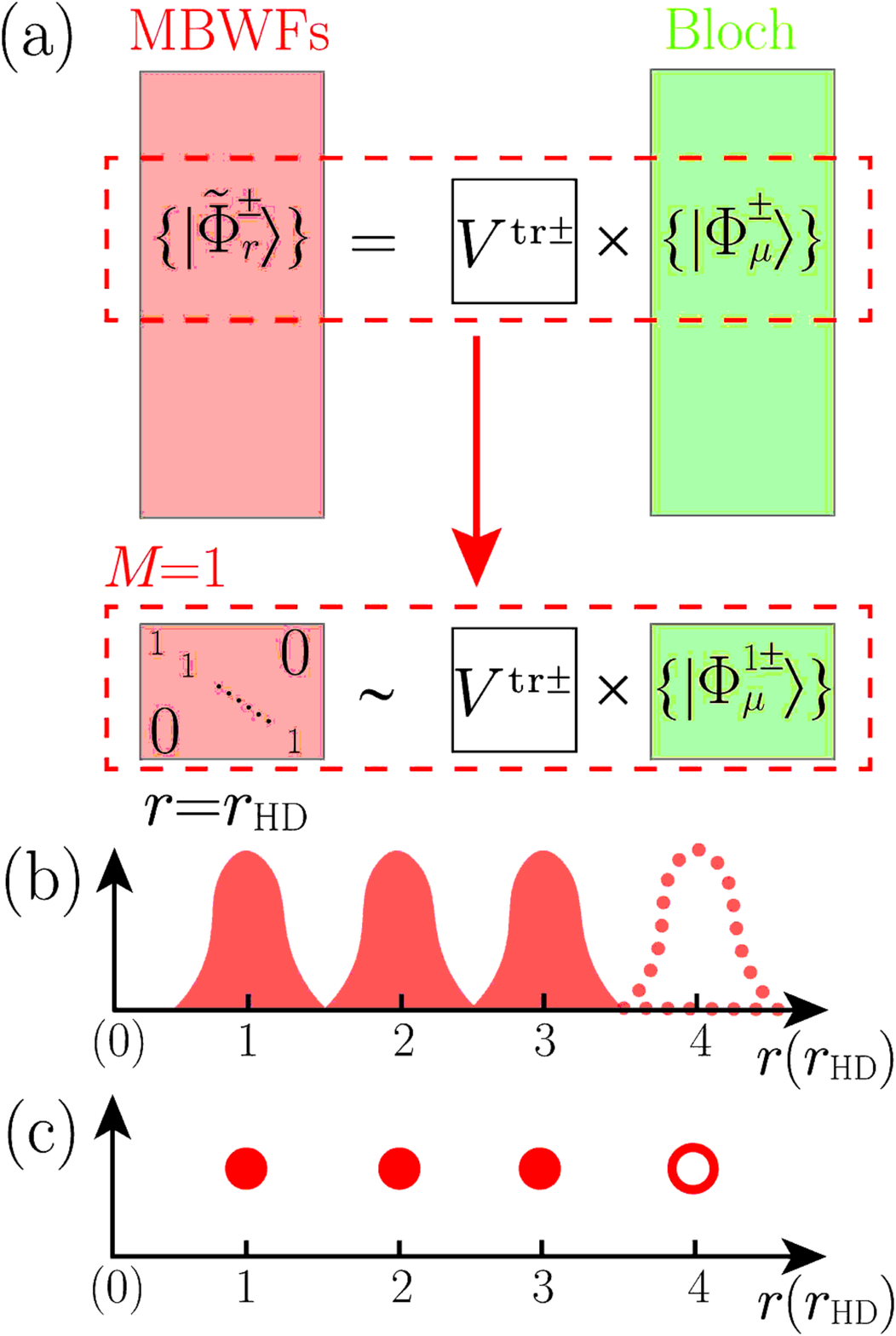}
\caption{
Schematics of the many-body Wannier functions (MBWFs) method for a charge model in this paper. 
(a) Unitary transformation, $V^{{\rm tr}p}$, from (Bloch) excited states (labeled by $\mu$) to MBWFs (labeled by $r$), where 
$p=+1$ ($-1$) denotes even- (odd-) parity. $V^{{\rm tr}p}$ in this paper should be selected to partially localize (Bloch) 
excited states as satisfying $r=r_{\rm HD}$. 
(b) Even-parity MBWFs of $N=6, M_{\rm max}=N/2=3$. Each filled diagram localizes around $r=r_{\rm HD}$. 
In the $M=1$ subspace of MBWFs, the size extension from $N$ to $N_{\rm ex}$ can naturally be achieved by adding 
$r_{\rm HD}=4,5,\cdots,I_{\rm ex}^{+}\equiv N_{\rm ex}/2$ 
MBWFs as schematically drown as a dotted diagram. 
(c) A certain physical quantity associated with even-parity excitations in the subspace labeled by $r$. This size extension is 
performed by extrapolation in the direction to increasing $r=r_{\rm HD}$. The filled circles and empty circle, which is one of the 
extrapolations, construct the physical quantity for $N_{\rm ex}=8$. 
}
\label{fig2}
\end{center}
\end{figure}
\par
\par
Now, defining an operator of translating one site to the right as $T_{\rm R}$ and in order to treat a simultaneous eigenstate of 
$T_{\rm R}$ and a parity inversion operator, $\mathcal{P}$, we introduce a bare basis, 
\begin{equation}
|r_{M}^{p}(K)\rangle
\equiv 
\frac{1}{ \sqrt{ \mathcal{N} } }
\sum_{l=0}^{N-1}
\cos(Kl)T_{\rm R}^{l}
(1+p\mathcal{P})|r_{M}\rangle, 
\label{eq14}
\end{equation}
where $\mathcal{N}$ is a normalization factor of $|r_{M}^{p}(K)\rangle$. $p=+1$ ($-1$) denotes an even- (odd-) parity state. 
In this paper, we basically assume the zero center-of-gravity momentum frame ($K=0$). The effect of finite $K=2\pi/N\ll 1$ is only 
treated for introducing $H_{\phi}^{({\rm C})}$ term into our effective models. Here, for $K\ll 1$, 
$T_{\rm R}|r_{M}^{p}(K)\rangle=\cos(K)|r_{M}^{p}(K)\rangle$ and $\mathcal{P}|r_{M}^{p}(K)\rangle=p|r_{M}^{p}(K)\rangle$ 
are derived, indeed. 
Extending our previous method\cite{CHRM} and as schematically explained in Fig.\ref{fig2} and its caption, 
we construct the many-body Wannier functions (MBWFs) and effective models to achieve the size extension 
from $N$ to $N_{\rm ex}\gg N$ in what follows. 
In the following subsections, we explain three key processes of our method with corresponding results. 
\subsection{First process: construction of MBWFs and size extension of $J^{({\rm C})}$ expectation values}
\label{sect2a}
%
%
\par
\begin{figure}[t]
\begin{center}
\includegraphics[width=8cm,keepaspectratio]{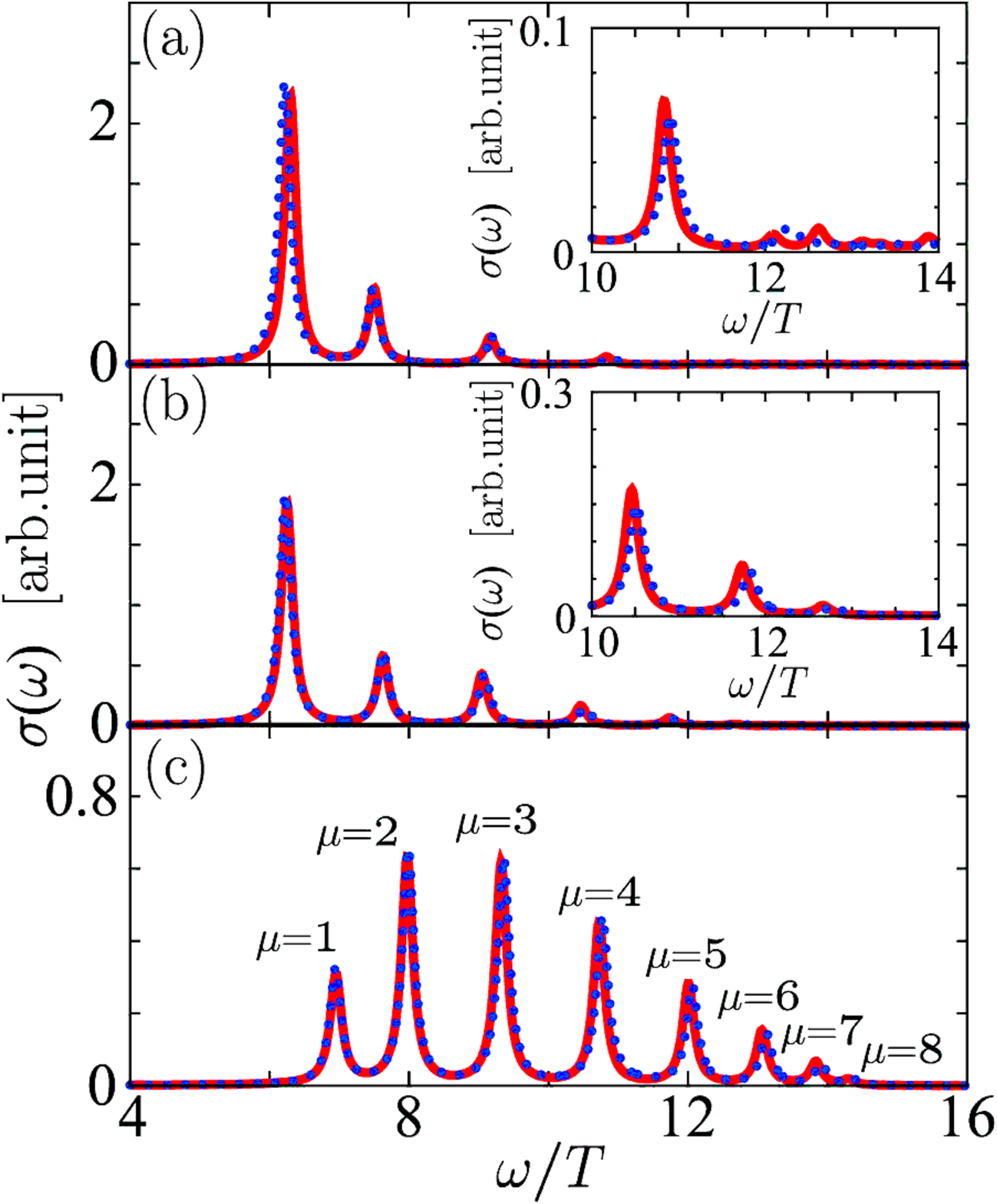}
\caption{Optical conductivity spectra of $N=18, \gamma=0.1T, \varepsilon=0$ for a charge model with 
$M_{\rm max}=N/2$ (solid lines) and half-filled 1D extended Hubbard model (dotted lines). 
(a) $V_{1}=V=2.8T, V_{\alpha\geq 2}=0$. (b) $V_{1}=V=2.4T, V_{2}=V/2, V_{3}=V/3, V_{\alpha\geq 4}=0$. 
Insets are optical spectra in large $\omega$ region. 
(c) $V_{\alpha\geq 1}=0$. $\mu$ denotes the labels of odd-parity photoexcited states. 
}
\label{fig3}
\end{center}
\end{figure}
\par
\par
\begin{figure}[t]
\begin{center}
\includegraphics[width=12cm,keepaspectratio]{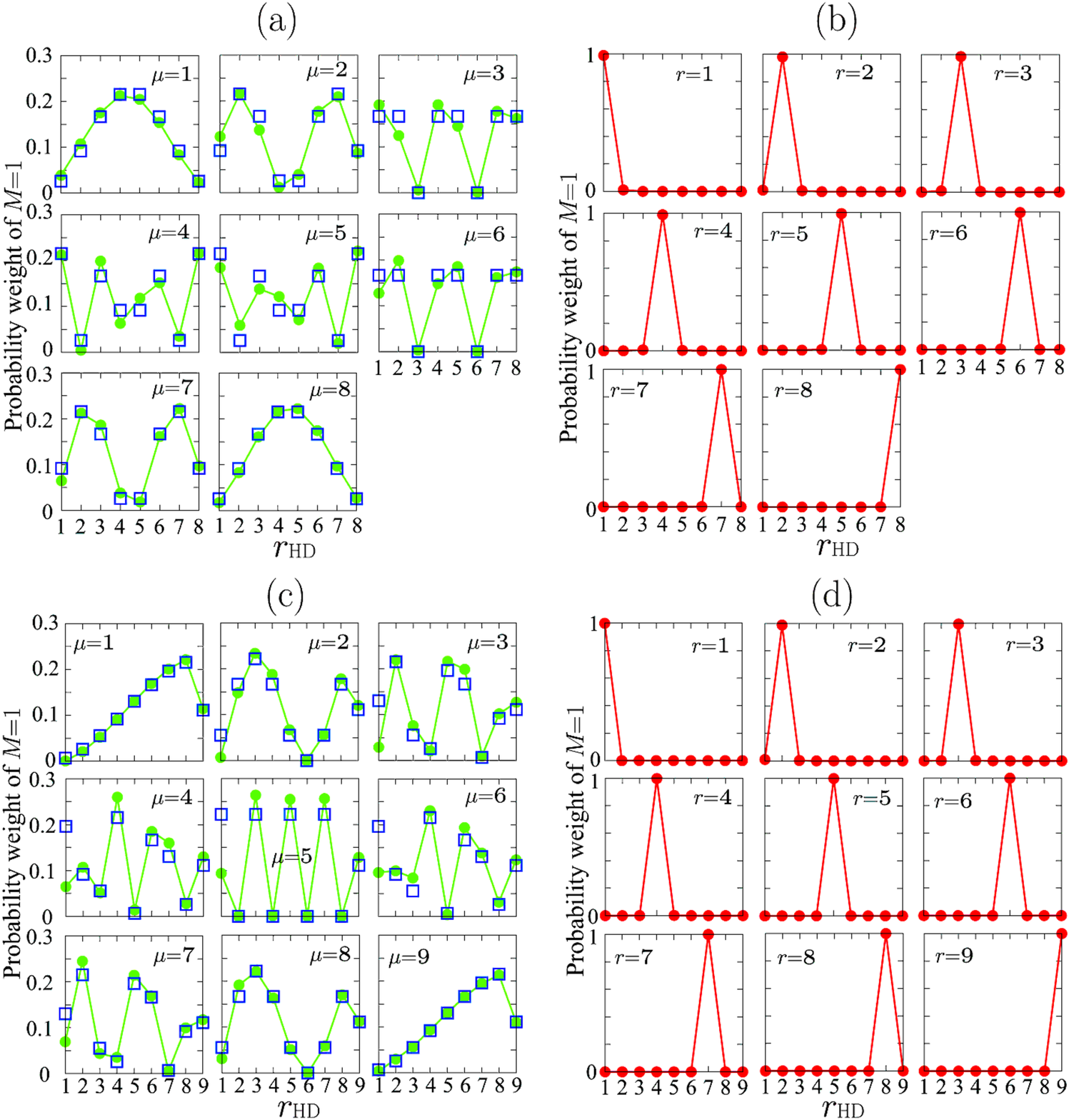}
\caption{
Probability weight in the $M=1$ subspace at $N=18$. Odd-parity Bloch states (a) and corresponding MBWFs (b). Even-parity 
Bloch states (c) and corresponding MBWFs (d). Filled circles are results with full charge fluctuations ($M_{\rm max}=N/2$) and 
empty squares are results of the HD model\cite{HDmodel}, briefly illustrated in Appendix \ref{AppdxA}. 
Solid lines are guides for eyes. 
}
\label{fig4}
\end{center}
\end{figure}
\par
\par
For finite $U$, the ground state, $|g\rangle$, is a Mott-insulator and has even-parity. Thus, all the excited states, 
corresponding to $J|g\rangle$, are odd-parity. As far as weak photoexcitations from $|g\rangle$ with sufficiently large $U$, 
a creation of a single holon-doublon (HD) pair is dominant. The number of such photoexcited (forbidden) states, $I$, is 
$N/2-1$ ($N/2$) with odd- (even-) parity, which are naturally classified by relative distance between a H and D, $r_{\rm HD}$. 
In this sense, the significant structure of $\sigma(\omega)$ in Eq.(\ref{eq9}) is typified by corresponding $N/2-1$ 
odd-parity eigenstates $|\Phi_{\mu}^{-}\rangle$ with their eigenenergies $E_{\mu}^{-}$. Thus, the approximation of 
\begin{equation}
\sigma(\omega) \sim \frac{\gamma}{N\omega}\sum_{\mu=1}^{I=N/2-1}
\frac{|\langle\Phi_{\mu}^{-}|J|g\rangle|^{2}}
{(\omega+E_{g}-E_{\mu}^{-})^{2} + \gamma^{2}}
\label{eq15}
\end{equation}
should be satisfied. 
However, as shown in Fig.\ref{fig3}, we cannot clearly distinguish significant $N/2-1$ peaks from the other peaks of 
$\sigma(\omega)$ with finite $V$. 
Furthermore, $\sigma(\omega)$ with finite $V$ of a charge model little discords from 
that of a 1D half-filled extended Hubbard model in large $\omega$ region. In contrast to this, for $V=0$, $\sigma(\omega)$ of 
a charge model well coincides with that of a 1D extended Hubbard model. Judging from this, we can regard all the excited states 
for a charge model as those for charge sectors of a 1D extended Hubbard model at least for low-energy excitations with $V=0$. 
\par
Due to above circumstance, for $V=0$ and $N=18$, we firstly calculate the ground state and all the low-energy excited states 
with full charge fluctuations ($M_{\rm max}=N/2$), 
\begin{equation}
H_{0}^{({\rm C})}|g^{({\rm C})}\rangle \equiv E_{g}^{({\rm C})}|g^{({\rm C})}\rangle, 
\label{eq16}
\end{equation}
\begin{equation}
H_{0}^{({\rm C})}|\Phi_{\mu}^{\pm}(0)\rangle \equiv E_{\mu}^{\pm}|\Phi_{\mu}^{\pm}(0)\rangle, 
\label{eq17}
\end{equation}
\begin{equation}
|\Phi_{\mu}^{\pm}(0)\rangle \equiv \sum_{r_{\rm M}^{\pm}} C_{\mu}(r_{\rm M}^{\pm})|r_{M}^{\pm}(0)\rangle, 
\label{eq18}
\end{equation}
by applying the Lanczos method\cite{LNCZ} within the framework of 
the exact diagonalization method ($E_{1}^{\pm}\leq\cdots\leq E_{I}^{\pm}$). For computational problem, $N=18$ is 
the maximum size of exact calculations in this paper. 
Those states are roughly interpreted as Bloch states of finite momentum $2\mu\pi/N$ ($(2\mu-1)\pi/N$) with 
$1\leq\mu\leq I=N/2-1$ ($I=N/2$) for odd- (even-) parity. 
\par
Next, we act unitary transformations, $V_{r\mu}^{{\rm tr}\pm}$, on 
$|\Phi_{\mu}^{\pm}(0)\rangle$ of Eq.(\ref{eq17}) in order to achieve one-to-one correspondence between $r$ and $r_{\rm HD}$ 
in the subspace of $M=1$, 
\begin{equation}
|\widetilde{\Phi}_{r}^{\pm}(0)\rangle\equiv \sum_{\mu} V_{r\mu}^{{\rm tr}\pm}|\Phi_{\mu}^{\pm}(0)\rangle. 
\label{eq19}
\end{equation}
Here, a subscript $r$ is in the range of $1\leq r\leq N/2$ for $|\widetilde{\Phi}_{r}^{+}(0)\rangle$ and 
$1\leq r\leq N/2-1$ for $|\widetilde{\Phi}_{r}^{-}(0)\rangle$, respectively.   
In this paper, we choose $V^{{\rm tr}\pm}$ as Eqs.(\ref{eqA3}), (\ref{eqA4}) because of the similarity of the probability weight 
of $|\Phi_{\mu}^{\pm}(0)\rangle$ in the subspace of $M=1$ between a charge model at $M_{\rm max}=N/2$ and the HD model, 
where the latter is validated in the limit of $U/T\rightarrow +\infty$\cite{HDmodel} (see Appendix \ref{AppdxA}). Defining 
$|\Phi_{\mu}^{1\pm}(0)\rangle$ as renormalized state vectors of $|\Phi_{\mu}^{\pm}(0)\rangle$ in the subspace of $M=1$, 
we show the similarity with respect to $|\langle\Phi_{\mu}^{1p}(0)|\Phi_{\mu}^{1p}(0)\rangle |^{2}$ in Figs.\ref{fig4} (a) for $p=-1$ 
and (c) for $p=+1$. Furthermore, in the same manner, we also illustrate the complete one-to-one connection between 
$r$ and $r_{\rm HD}$ of MBWFs by calculating $|\langle\widetilde{\Phi}_{r}^{1p}(0)|\widetilde{\Phi}_{r}^{1p}(0)\rangle |^{2}$ 
as shown in Figs.\ref{fig4} (b) for $p=-1$ and (d) for $p=+1$. This complete localization of MBWFs regarding $r_{\rm HD}$ 
validates the extrapolation procedure with size extension in what follows. 
Here, the $N/2$ ($N/2-1$) vector components in the $M=1$ subspace of $|\Phi_{\mu}^{+}(0)\rangle$ ($|\Phi_{\mu}^{-}(0)\rangle$) 
in Eq.(\ref{eq17}) account for approximately 63-68 (67-73)\% of all the vector components 
as opposed to 100\% of the HD model. 
The difference about 30\% is related to the many-body effects originating from the creations or annihilations of HD pairs. 
Then, we appropriately include them into our effective models with MBWFs as introduced in the next subsection. 
\par
In the end of this subsection, we calculate expectation values of $J^{({\rm C})}$ in Eq.(\ref{eq11}) by defining 
\begin{equation}
J_{rg}^{\pm}\equiv \frac{\langle\widetilde{\Phi}_{r}^{\pm}(0)|J^{({\rm C})}|g^{({\rm C})}\rangle}{\sqrt{N}}. 
\label{eq20}
\end{equation}
Because $|g^{({\rm C})}\rangle$ in Eq.(\ref{eq16}) is even-parity, $J_{rg}^{+}=0$ for any $r$. 
In contrast, $J_{rg}^{-}$ ($1\leq r\leq 8$) at $N=18$ have finite values as shown in Fig.\ref{fig5}(a). 
However, as clearly seen in that figure, we can presume that $J_{8g}^{-}$ converges to almost zero. In this sense, following 
extrapolation is permitted as one of size extended formulations of expectation values of $J^{({\rm C})}$ at $N_{\rm ex}$. 
\begin{align}
(J_{\rm ex})_{rg}^{+} &= 0 \quad (1\leq r\leq I_{\rm ex}^{+}=N_{\rm ex}/2), 
\label{eq21} \\
(J_{\rm ex})_{rg}^{-} &\equiv 
\begin{cases}
  J_{rg}^{-}  & (1\leq r\leq 8) \\
       0       & (9\leq r\leq I_{\rm ex}^{-}=N_{\rm ex}/2-1)
\end{cases}. 
\label{eq22}
\end{align}
\subsection{Second process: construction and size extension of effective models with perturbative $H_{V}^{({\rm C})}$}
\label{sect2b}
%
%
\par
\begin{figure}[t]
\begin{center}
\includegraphics[width=12cm,keepaspectratio]{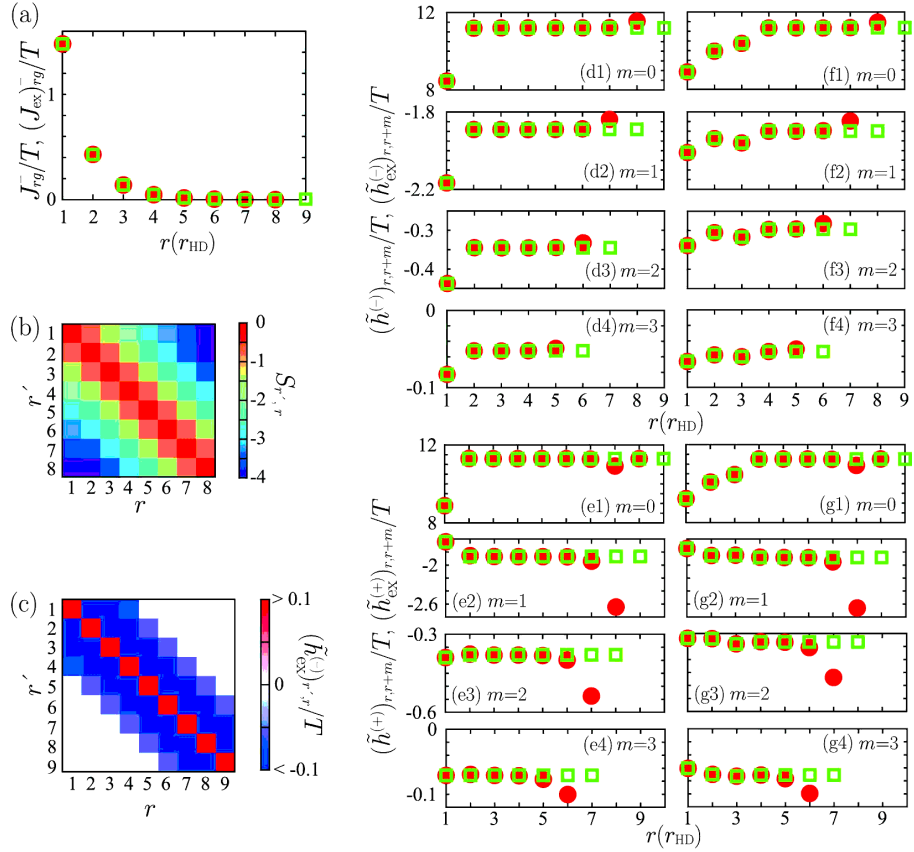}
\caption{
(a) Expectation values of the charge-current operator at $N=18$ (filled circles) and their size extended values to 
$N_{\rm ex}=20$ (empty squares). 
(b), (c), (d1)-(d4), (e1)-(e4), (f1)-(f4), and (g1)-(g4) are effective models at $N=18$ (filled circles) and their size extended values 
from $N=18$ to $N_{\rm ex}=20$ (empty squares). 
(b) $S_{r^{\prime},r}\equiv \log_{10}(|(\widetilde{h}^{-})_{r^{\prime},r}|/\widetilde{h}_{\rm max}^{-})$ with 
$V_{1}=V=2.8T, V_{\alpha\geq 2}=0$. 
$\widetilde{h}_{\rm max}^{-}\equiv \max(|(\widetilde{h}^{-})_{r^{\prime},r}|)$. Whole structure of $S_{r^{\prime},r}$ is 
almost the same as all the other effective models calculated in this paper. 
(c) Size extended odd-parity effective model, $\widetilde{h}_{\rm ex}^{-}$, at $N_{\rm ex}=20$ 
with $V_{1}=V=2.8T, V_{\alpha\geq 2}=0$. All the $S_{r^{\prime},r}<10^{-3}$ terms are ignored in this model. 
(d1-d4) Results with odd-parity for $V_{1}=V=2.8T, V_{\alpha\geq 2}=0$. 
(f1-f4) Results with odd-parity for $V_{1}=V=2.4T, V_{2}=V/2, V_{3}=V/3, V_{\alpha\geq 4}=0$. 
(e1-e4) Results with even-parity for $V_{1}=V=2.8T, V_{\alpha\geq 2}=0$. 
(g1-g4) Results with even-parity for $V_{1}=V=2.4T, V_{2}=V/2, V_{3}=V/3, V_{\alpha\geq 4}=0$.
}
\label{fig5}
\end{center}
\end{figure}
\par
%
%
%
%
\par
\begin{figure}[t]
\begin{center}
\includegraphics[width=8cm,keepaspectratio]{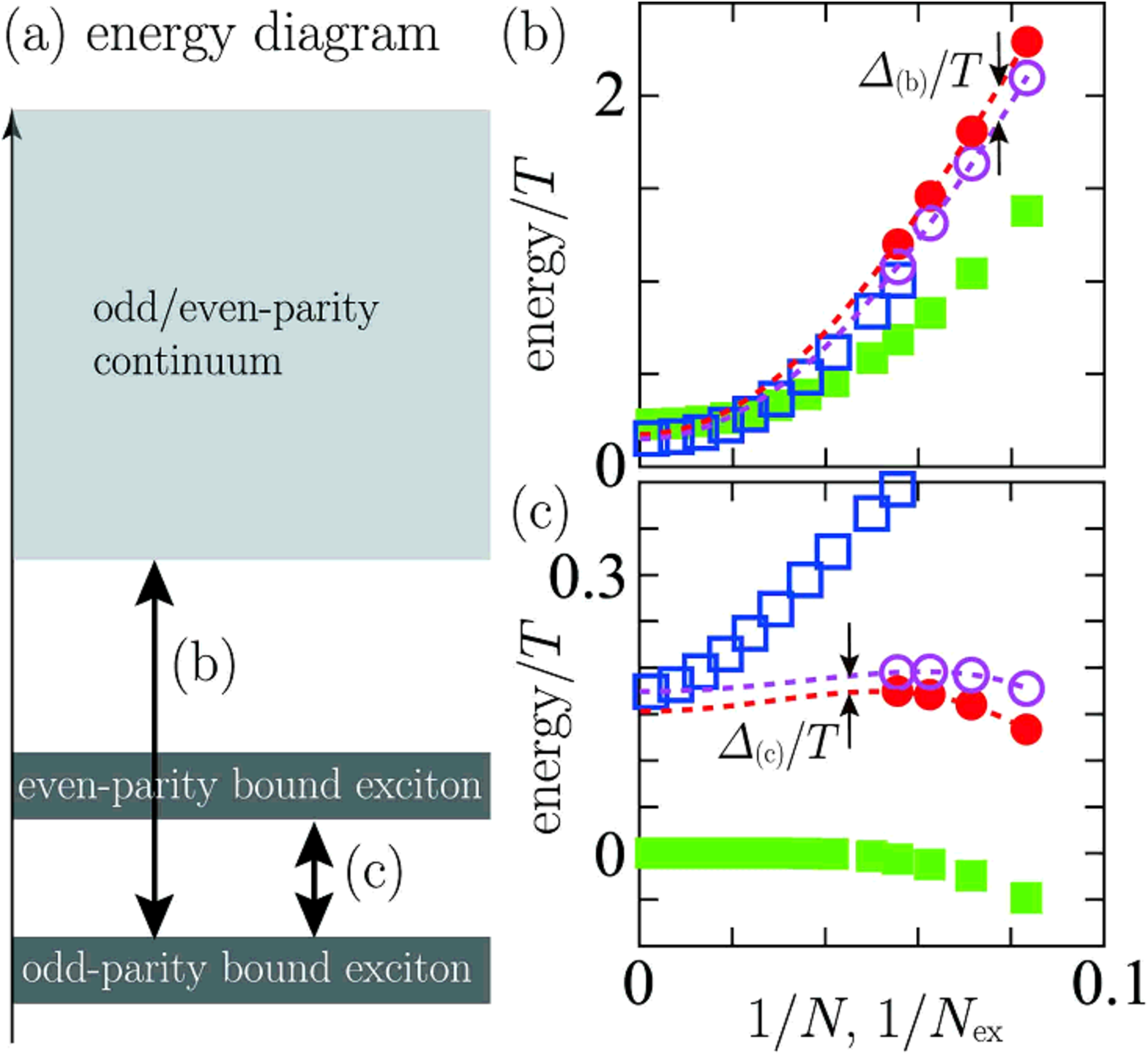}
\caption{
(a) Schematic of an excitonic energy structure of a Mott insulator in the region of low-lying excitation energies. 
(b), (c) Finite size effects of the relative energies corresponding to the notations in (a) for $V_{1}=V=2.8T$, $V_{\alpha\geq 2}=0$. 
The values of filled and empty circles are calculated by using the eigenenergies of the exact diagonalization method for 
$M_{\rm max}=N/2$ and of $\widetilde{h}^{\pm}$ in Eq.(\ref{eq23}) with $N=12$, 14, 16, 18, respectively. 
Both the filled and empty circles are fitted with the power series of $1/N$ as shown with the broken lines. 
Eigenenergies of the size-extended effective model in Eqs.(\ref{eq24})-(\ref{eq29}) with 
$N_{\rm ex}=18$, 20, 24, 28, 34, 42, 54, 76, 128, 400 give the values of empty squares. 
Those of the HD model\cite{HDmodel} for $N=12$, 14, 16, 18, 20, 24, 28, 34, 42, 54, 76, 128, 400 correspond to the filled squares. 
For $N, N_{\rm ex}>20$,  we choose $N$ and $N_{\rm ex}$ values to keep as equal intervals in the horizontal axis as possible. 
In (c), the HD model results never quantitatively converge to a plausible value of extrapolating exact results (filled circles) 
in the limit of $N\rightarrow +\infty$.  
}
\label{fig6}
\end{center}
\end{figure}
\par
\par
In this subsection, we treat a long-range Coulomb interaction term, $H_{V}^{({\rm C})}$ (see Eqs.(\ref{eq3}) and (\ref{eq10})), 
as a perturbation and calculate the effective models of $H^{({\rm C})}$ in Eq.(\ref{eq13}) 
by utilizing previous MBWFs, $|\widetilde{\Phi}_{r}^{\pm}(0)\rangle$ in Eq.(\ref{eq19}). 
As an example, we can introduce an effective model by calculating following matrix elements\cite{Comment0}. 
\begin{align}
\left(
\widetilde{h}^{\pm}
\right)_{r^{\prime},r} &\equiv 
\langle\widetilde{\Phi}_{r^{\prime}}^{\pm}(0)| H^{({\rm C})} - E_{gV}^{({\rm C})} |\widetilde{\Phi}_{r}^{\pm}(0)\rangle 
\nonumber 
\\
&= \langle\widetilde{\Phi}_{r^{\prime}}^{\pm}(0)| H_{0}^{({\rm C})} +  H_{V}^{({\rm C})} - E_{gV}^{({\rm C})} 
|\widetilde{\Phi}_{r}^{\pm}(0)\rangle, 
\label{eq23}
\end{align}
where $E_{gV}^{({\rm C})}\equiv E_{g}^{({\rm C})}+\langle g^{({\rm C})}|H_{V}^{({\rm C})}|g^{({\rm C})}\rangle$ for 
$N=18$ and $M_{\rm max}=N/2$. 
In this paper, the size extensions of $\widetilde{h}^{\pm}$ are achieved by extending our previous method\cite{CHRM}. 
Exemplifying the case of $\widetilde{h}^{-}$ with $V_{1}\equiv V, V_{\alpha\geq 2} = 0$, we explain how to access 
the size extended effective models from $N=18$ of Eq.(\ref{eq23}) to $N_{\rm ex}\gg N$. 
\par
Firstly, we focus on the order of all the absolute values of the matrix elements $|(\widetilde{h}^{-})_{r^{\prime},r}|$. Defining the 
maximum value, $\widetilde{h}_{\rm max}^{-}\equiv \max(|(\widetilde{h}^{-})_{r^{\prime},r}|)$, we introduce 
$S_{r^{\prime},r}\equiv \log_{10}(|(\widetilde{h}^{-})_{r^{\prime},r}|/\widetilde{h}_{\rm max}^{-})$ for convenience. 
$S_{r^{\prime},r}$ for $N=18$ is shown in Fig.\ref{fig5}(b). This diagonal structure is naturally understood because 
$(\widetilde{h}^{-})_{r,r+m}$ is roughly related to the $m$th nearest neighbor transfer integral originating from 
the $m$th order perturbative expansion of $T/U$. 
Then, we ignore all the matrix elements of $S_{r^{\prime},r}<10^{-3}$ to construct the corresponding size extended effective 
model, $\widetilde{h}_{\rm ex}^{-}$, in this paper. 
Because the diagonal structure and the order of $S_{r^{\prime},r}$ for all the effective models calculated in this paper 
are almost the same as the case of above $\widetilde{h}^{-}$ with $V_{1}\equiv V, V_{\alpha\geq 2} = 0$, we similarly 
neglect $S_{r^{\prime},r}<10^{-3}$ matrix elements for all the other effective models. 
Note that one can, of course, remain more matrix elements to obtain more accurate effective models. However, this increases 
the difficulty of extrapolating $\widetilde{h}^{\pm}$ to a size extended one, $\widetilde{h}_{\rm ex}^{\pm}$, owing to reduction in 
the number of the matrix elements, which are essential for the extrapolation toward a size extension. 
\par
Next, we extract all the matrix elements of $S_{r^{\prime},r}>10^{-3}$ at $N=18$, $\widetilde{h}_{r,r+m}^{-}$ ($0\leq m\leq 3$), 
as filled circles shown in Figs.\ref{fig5}(d1)-(d4). There are two significant features at the both edges, namely, 
$\widetilde{h}_{1,1+m}^{-}$ and $\widetilde{h}_{8-m,8}^{-}$. The former just corresponds to an attractive Coulomb 
interaction concerning an excitonic state at $r(r_{\rm HD})=1$. This should be included in the size extended model. 
The latter can be interpreted as the finite size effect at the edge, $r(r_{\rm HD})=8$. Because we now consider the size extension 
in the direction to increasing $r(r_{\rm HD})$ as $r(r_{\rm HD})=9,\cdots,I_{\rm ex}^{-}=N_{\rm ex}/2-1$, such edge effect can be 
neglected for sufficiently large $N_{\rm ex}$. Regarding the rest of $\widetilde{h}_{r,r+m}^{-}$, all the calculated values for fixed $m$ 
are subequal. Then, for each $m$, we simply substitute the arithmetic mean value of 
$\widetilde{h}_{r,r+m}^{-}$ ($2\leq r\leq 7-m$) into corresponding matrix elements of a size extended effective model. 
Note that this approximation is also applied in our previous paper\cite{CHRM}. 
Consequently, our effective model with size extension, $\widetilde{h}_{\rm ex}^{-}$, is written in the following matrix elements. 
Namely, for $0\leq m\leq 3$, 
\begin{equation}
\left(
\widetilde{h}_{\rm ex}^{-}
\right)_{1,1+m} =
\left(
\widetilde{h}_{\rm ex}^{-}
\right)_{1+m,1} \equiv 
\left(
\widetilde{h}^{-}
\right)_{1,1+m}, 
\label{eq24}
\end{equation}
for $0\leq m\leq 3$ and $2\leq r\leq I_{\rm ex}^{-}-m$, 
\begin{equation}
\left(
\widetilde{h}_{\rm ex}^{-}
\right)_{r,r+m} = 
\left(
\widetilde{h}_{\rm ex}^{-}
\right)_{r+m,r} \equiv \frac{1}{6-m}\sum_{l=2}^{7-m}
\left(
\widetilde{h}^{-}
\right)_{l,l+m},  
\label{eq25}
\end{equation}
and for $m\geq 4$, 
\begin{equation}
\left(
\widetilde{h}_{\rm ex}^{-}
\right)_{r,r+m} = 
\left(
\widetilde{h}_{\rm ex}^{-}
\right)_{r+m,r} \equiv 0.  
\label{eq26}
\end{equation}
In one instance, $\widetilde{h}_{\rm ex}^{-}$ at $N_{\rm ex}=20$ is shown in Fig.\ref{fig5}(c) and 
as empty squares in Figs.\ref{fig5}(d1)-(d4). 
\par
Above situation is almost the same as the even-parity case with $V_{1}\equiv V, V_{\alpha\geq 2} = 0$ except for 
increasing one more edge effect at $r(r_{\rm HD})=7$ (see filled circles in Figs.\ref{fig5}(e1)-(e4)). 
This is roughly attributed to discontinuous structure of $V_{r\mu}^{{\rm tr}+}$ clearly seen in Eq.(\ref{eqA3}). 
However, such edge effect should be also removed for adequately large $N_{\rm ex}$ as discussed above. 
Thus, as one of the size extended effective models, $\widetilde{h}_{\rm ex}^{+}$, 
shown for $N_{\rm ex}=20$ in Figs.\ref{fig5}(e1)-(e4), 
is also written in the following matrix elements as, for $0\leq m\leq 3$ and $2\leq r\leq I_{\rm ex}^{+}-m$ 
($I_{\rm ex}^{+}\equiv N_{\rm ex}/2$), 
\begin{equation}
\left(
\widetilde{h}_{\rm ex}^{+}
\right)_{1,1+m} =
\left(
\widetilde{h}_{\rm ex}^{+}
\right)_{1+m,1} \equiv 
\left(
\widetilde{h}^{+}
\right)_{1,1+m}, 
\label{eq27}
\end{equation}
\begin{equation}
\left(
\widetilde{h}_{\rm ex}^{+}
\right)_{r,r+m} = 
\left(
\widetilde{h}_{\rm ex}^{+}
\right)_{r+m,r} \equiv \frac{1}{6-m}\sum_{l=2}^{7-m}
\left(
\widetilde{h}^{+}
\right)_{l,l+m},  
\label{eq28}
\end{equation}
and for $m\geq 4$, 
\begin{equation}
\left(
\widetilde{h}_{\rm ex}^{+}
\right)_{r,r+m} = 
\left(
\widetilde{h}_{\rm ex}^{+}
\right)_{r+m,r} \equiv 0.
\label{eq29}
\end{equation}
\par
The size extension of $\widetilde{h}^{\pm}$ in the case of $V_{\alpha}\equiv V/\alpha \; (\alpha=1,2,3), V_{\alpha\geq 4} = 0$ 
is achieved in almost the same way as already mentioned. The specific difference is $\alpha$ dependency of 
$(\widetilde{h}_{\rm ex}^{\pm})_{\alpha,\alpha+m}$ for $\alpha=1,2,3$. This is simply because exciton effects associated with 
$V_{\alpha}\equiv V/\alpha \; (\alpha=1,2,3)$ terms and same structure is also seen in corresponding matrix elements of the 
holon-doublon (HD) model for $m=0$ (see diagonal elements in Eqs.(\ref{eqA1}) and (\ref{eqA2})). 
Then, the size extended effective models $\widetilde{h}_{\rm ex}^{\pm}$ in this paper, 
shown for $N_{\rm ex}=20$ in Figs.\ref{fig5}(f1)-(f4) and (g1)-(g4), are similarly defined as, for $0\leq m\leq 3$ and 
$4\leq r\leq I_{\rm ex}^{\pm}-m$, 
\begin{equation}
\left(
\widetilde{h}_{\rm ex}^{\pm}
\right)_{\alpha,\alpha+m} =
\left(
\widetilde{h}_{\rm ex}^{\pm}
\right)_{\alpha+m,\alpha} \equiv 
\left(
\widetilde{h}^{\pm}
\right)_{\alpha,\alpha+m}, 
\label{eq30}
\end{equation}
\begin{equation}
\left(
\widetilde{h}_{\rm ex}^{\pm}
\right)_{r,r+m} = 
\left(
\widetilde{h}_{\rm ex}^{\pm}
\right)_{r+m,r} \equiv \frac{1}{4-m}\sum_{l=4}^{7-m}
\left(
\widetilde{h}^{\pm}
\right)_{l,l+m},  
\label{eq31}
\end{equation}
and for $m\geq 4$, 
\begin{equation}
\left(
\widetilde{h}_{\rm ex}^{\pm}
\right)_{r,r+m} = 
\left(
\widetilde{h}_{\rm ex}^{\pm}
\right)_{r+m,r} \equiv 0.  
\label{eq32}
\end{equation}
\par
In the end of this subsection, as exhibited in Fig.\ref{fig6}, we check the finite size effects of above effective models comparing 
with those of other models by exemplifying the case of $V_{1}=V$, $V_{\alpha\geq 2}=0$. 
Using above effective models, the excitonic energy structure is theoretically determined by solving the eigenenergies of 
$\widetilde{h}^{\pm}$ in Eq.(\ref{eq23}) and $\widetilde{h}_{\rm ex}^{\pm}$ in Eqs.(\ref{eq24})-(\ref{eq29}), which are shown as 
empty circles and empty squares, respectively, in Figs.\ref{fig6}(b), (c). 
Firstly, the former structure is in good agreement with the structure of filled circles in Figs.\ref{fig6}(b), (c), which are calculated 
by the precise eigenenergies of exactly diagonalizing $H_{0}^{({\rm C})}+H_{V}^{({\rm C})}$ in Eq.(\ref{eq13}) at $M_{\rm max}=N/2$ 
within the framework of the Lanczos method at finite sizes. 
In contrast, the latter structure is, of course, affected by finite size effects for small $N_{\rm ex}$ to some extent due to 
above explained approximations of constructing the size-extended effective models. 
However, in the limit of $N_{\rm ex}\rightarrow +\infty$, the latter seems to quantitatively converge to a plausible value of 
extrapolating exact results (filled circles) for $N\rightarrow +\infty$. 
In this regard, we can confirm the sufficient inclusion of the many-body effects in above constructed effective models, 
$\widetilde{h}_{\rm ex}^{\pm}$, with $N_{\rm ex}\gg N=18$. 
Finally, we evaluate the order of the error of the eigenenergies in our effective models caused by above-mentioned perturbative 
treatment of $H_{V}^{({\rm C})}$. Defining $\Delta_{({\rm b})}$ ($\Delta_{({\rm c})}$) as the difference between two fitting 
functions, which are shown as the broken lines in Fig.\ref{fig6}(b) ((c)), and deducing from $\Delta_{({\rm b}), ({\rm c})}\sim 0.02T$ 
for $N\rightarrow +\infty$, the order of the error is estimated to be $10^{-2}T$. This is in the order of 1 meV for 
ET-F$_{2}$TCNQ and practically negligible. 
\subsection{Third process: introduction of $H_{\phi}^{({\rm C})}$ to an effective model and its size extension}
\label{sect2c}
\par
In this subsection, we directly treat $H_{\phi}^{({\rm C})}$ term (see Eqs.(\ref{eq4}) and (\ref{eq10})) with the approximation of 
the finite momentum, $K\equiv 2\pi/N\ll 1$. Namely, we approximately define the MBWFs with the finite momentum $K$ as 
\begin{equation}
|\widetilde{\Phi}_{r}^{\pm}(K)\rangle \equiv \sum_{r_{\rm M}^{\pm}} W_{r}(r_{\rm M}^{\pm})|r_{M}^{\pm}(K)\rangle, 
\label{eq33}
\end{equation}
\begin{equation}
|\widetilde{\Phi}_{r}^{\pm}(0)\rangle \equiv \sum_{r_{\rm M}^{\pm}} W_{r}(r_{\rm M}^{\pm})|r_{M}^{\pm}(0)\rangle, 
\label{eq34}
\end{equation}
where $|\widetilde{\Phi}_{r}^{\pm}(0)\rangle$ is in Eq.(\ref{eq19}). 
In a strict sense, many-body effects of $H_{\phi}^{({\rm C})}$ should be considered for producing our effective models. 
However, in this paper, the $H_{\phi}^{({\rm C})}$ term is only required for mixing the effective models $\widetilde{h}_{\rm ex}^{+}$ 
and $\widetilde{h}_{\rm ex}^{-}$, calculated in the previous subsection. 
Regarding this, we ignore the many-body effects and only consider the $M=1$ subspace of $H_{\phi}^{({\rm C})}$ to introduce 
the effective models of $H_{\phi}^{({\rm C})}$ in this paper. 
Then, due to $|\widetilde{\Phi}_{r}^{\pm}(K)\rangle=|r_{M}^{\pm}(K)\rangle$ by definition, 
\begin{align}
\left(
\widetilde{h}^{\phi}
\right)_{r^{\prime},r} &\equiv 
\frac{ \langle r_{1}^{\prime +}(K)| H^{({\rm C})} |r_{1}^{-}(0)\rangle 
+ \langle r_{1}^{\prime +}(0)| H^{({\rm C})} |r_{1}^{-}(K)\rangle  }{2}
\nonumber 
\\
&= \frac{ \langle r_{1}^{\prime +}(K)| H_{\phi}^{({\rm C})} |r_{1}^{-}(0)\rangle 
+ \langle r_{1}^{\prime +}(0)| H_{\phi}^{({\rm C})} |r_{1}^{-}(K)\rangle  }{2}
\nonumber 
\\
&= 2\phi_{0}
\sin\left(
\frac{Kr}{2}
\right)
\cos\left(
\frac{K(r+2)}{2}
\right)\delta_{r^{\prime},r}, 
\label{eq35}
\end{align}
is derived after little cumbersome calculations for $1\leq r^{\prime}\leq N/2$ and $1\leq r\leq N/2-1$. 
This can be easily augmented in the direction to increasing $r^{\prime}, r$ and one finds 
$\left(
\widetilde{h}_{\rm ex}^{\phi}
\right)_{r^{\prime},r}\equiv
\left(
\widetilde{h}^{\phi}
\right)_{r^{\prime},r} 
(1\leq r^{\prime}\leq N_{\rm ex}/2, 1\leq r\leq N_{\rm ex}/2-1)$. 
We confirm the validity of this approximation at least for finite size exact calculations with $V_{1}=V, V_{\alpha\geq 2}=0$ case 
in Fig.\ref{fig9}(b) of Appendix \ref{AppdxB}. However, to discuss whether the many-body effects of $H_{\phi}^{({\rm C})}$ 
at large system size are crucial or not is positioned as one of our future works. 
\par
To summarize above all the results, the effective model of the full form of $H^{({\rm C})}$ in Eq.(\ref{eq13}) for $N=18$ is defined as, 
\begin{align}
\widetilde{h} &\equiv 
\begin{bmatrix}
\widetilde{h}^{+} & \widetilde{h}^{\phi} \\
\widetilde{h}^{\phi\dagger} & \widetilde{h}^{-}
\end{bmatrix}, 
\label{eq36}
\end{align}
and its size extension to $N_{\rm ex}$ is also defined as, 
\begin{align}
\widetilde{h}_{\rm ex} &\equiv 
\begin{bmatrix}
\widetilde{h}_{\rm ex}^{+} & \widetilde{h}_{\rm ex}^{\phi} \\
\widetilde{h}_{\rm ex}^{\phi\dagger} & \widetilde{h}_{\rm ex}^{-}
\end{bmatrix}. 
\label{eq37}
\end{align}
Introducing the eigenstates and eigenvalues of $\widetilde{h}_{\rm ex}$ as 
\begin{equation}
\widetilde{h}_{\rm ex}|\psi_{\lambda}\rangle\equiv \omega_{\lambda}|\psi_{\lambda}\rangle, 
\label{eq38}
\end{equation}
the optical conductivity spectrum at $N_{\rm ex}$ is written as 
\begin{equation}
\sigma(\omega) = \frac{\gamma}{\omega}
\sum_{\lambda=1}^{N_{\rm ex}-1}
\frac{|\langle\psi_{\lambda}|J_{\rm ex}\rangle|^{2}}
{(\omega-\omega_{\lambda})^{2} + \gamma^{2}}, 
\label{eq39}
\end{equation}
where 
\begin{equation}
|J_{\rm ex}\rangle \equiv \left[(J_{\rm ex})_{1g}^{+}, \cdots, (J_{\rm ex})_{I_{\rm ex}^{+}g}^{+}, 
(J_{\rm ex})_{1g}^{-}, \cdots, (J_{\rm ex})_{I_{\rm ex}^{-}g}^{-} \right]^{\dagger}. 
\label{eq40}
\end{equation}
Here, all the vector elements of $|J_{\rm ex}\rangle$ are in Eqs.(\ref{eq21}) and (\ref{eq22}). Lastly, 
we introduce the change of optical spectra between with and without a modulated electric field, 
\begin{equation}
\Delta\sigma(\omega) \equiv \sigma(\omega)|_{\varepsilon\neq 0} - \sigma(\omega)|_{\varepsilon=0}, 
\label{eq41}
\end{equation}
and $\sigma_{\rm max}\equiv {\rm max}(\sigma(\omega)|_{\varepsilon=0})$. 
In a strict sense, the lifetimes of the eigenstates in Eq.(\ref{eq38}) with finite $\varepsilon$ differ from those without 
$\varepsilon$. To simply incorporate this into our theoretical calculations, we distinguish between the broadenings of 
calculating $\Delta\sigma(\omega)$ and those of calculating only $\sigma(\omega)|_{\varepsilon=0}$ itself. 
Hereafter, our calculations 
employ $N_{\rm ex}=400$, which is a sufficiently large size to provide $\sigma(\omega)|_{\varepsilon=0}$ 
with negligible finite size effects. 
%
%
%
\section{Optical spectra and excitonic energy structure}
\label{sect3}
\par
\begin{figure}[t]
\begin{center}
\includegraphics[width=10cm,keepaspectratio]{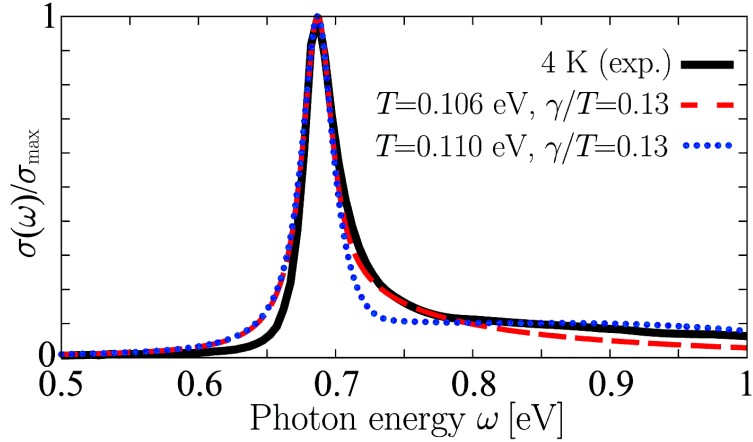}
\caption{Optical conductivity spectra with $\varepsilon=0$. The solid line is the experimental spectrum of ET-F$_{2}$TCNQ at 4 K, 
which has $\sigma_{\rm max}=2586$ S/cm at $\omega=0.687$ eV. This spectrum is obtained by the Kramers-Kronig transformation 
of the reflectivity spectrum with the electric fields of lights parallel to the 1D molecular stack. The experimental method of the 
low-temperature reflectivity measurement was reported in Ref.\cite{4Kexp}. 
The dashed (dotted) line represents the MBWFs spectrum with size extension from 
$N=18$ to $N_{\rm ex}=400$ and $V_{1}=V=2.8T, V_{\alpha\geq 2}=0$ ($V_{1}=V=2.4T, V_{2}=V/2, V_{3}=V/3, V_{\alpha\geq 4}=0$). 
}
\label{fig7}
\end{center}
\end{figure}
\par
\par
\begin{figure}[t]
\begin{center}
\includegraphics[width=\linewidth,keepaspectratio]{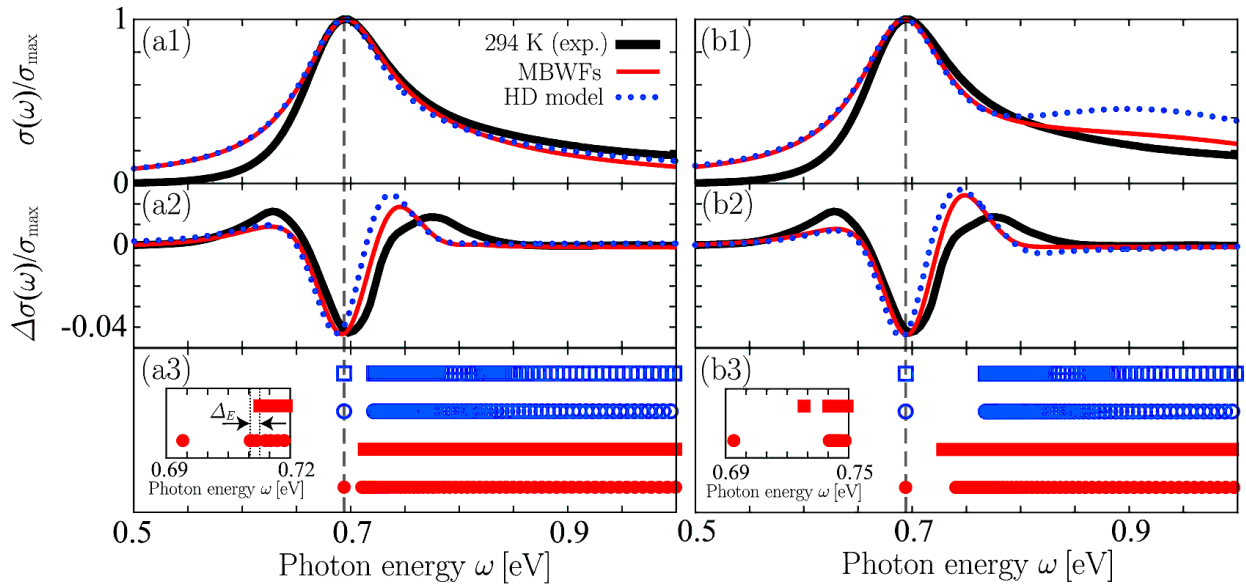}
\caption{Optical conductivity spectra with and without a modulated electric field. The solid thick lines are experimental data 
of ET-F$_{2}$TCNQ at 294 K, where $\sigma_{\rm max}=1119$ S/cm is observed at $\omega=0.694$ eV (the vertical dashed line). 
The solid thin lines are MBWFs spectra with size extension from $N=18$ to $N_{\rm ex}=400$. 
The dotted lines are spectra of the HD model with $N=400$. 
(a1-a3) $V_{1}=V=2.8T, V_{\alpha\geq 2}=0$, and $\varepsilon=1.31\times 10^{-2}$ ($1.25\times 10^{-2}$) for MBWFs (HD model). 
This corresponds to $E_{\rm amp}=152$ (163) kV/cm, due to $T=0.107$ (0.120) eV for MBWFs (HD model). 
$\gamma/T=0.45$ (0.40) in (a1) and $\gamma/T=0.55$ (0.50) in (a2) for MBWFs (HD model). 
(b1-b3) $V_{1}=V=2.4T, V_{2}=V/2, V_{3}=V/3, V_{\alpha\geq 4}=0$, and 
$\varepsilon=9.53\times 10^{-3}$ ($7.90\times 10^{-3}$) for MBWFs (HD model). 
This corresponds to $E_{\rm amp}=115$ (110) kV/cm, due to $T=0.111$ (0.128) eV for MBWFs (HD model). 
$\gamma/T=0.45$ (0.39) in (b1) and $\gamma/T=0.55$ (0.44) in (b2) for MBWFs (HD model). 
(a3), (b3) are eigenenergies measured from the ground state energy. Filled circles (squares) represent eigenenergies of MBWFs with 
odd- (even-) parity. Empty circles (squares) represent eigenenergies of the HD model with odd- (even-) parity. 
Insets are magnified low-lying energy diagrams of MBWFs. 
}
\label{fig8}
\end{center}
\end{figure}
\par
\par
Firstly, we determine long range Coulomb interaction strengths, $V$s, by reproducing the optical conductivity spectrum 
with $\varepsilon=0$ of ET-F$_{2}$TCNQ, $\sigma(\omega)$, newly measured at 4 K. 
We consider that this spectrum contains pure electronic photoexcitations with almost negligible electron-phonon couplings. 
As shown in Fig.\ref{fig7}, we found the best $V$s for 
\begin{enumerate}
 \item $V_{1}=V=2.8T, V_{\alpha\geq 2}=0$, 
 \item $V_{1}=V=2.4T, V_{2}=V/2, V_{3}=V/3, V_{\alpha\geq 4}=0$ 
\end{enumerate}
(see Eqs.(\ref{eq6}) and (\ref{eq7})). 
\par
Next, tuning artificial broadenings, $\gamma$, we reproduce 
observed $\sigma(\omega)$ of ET-F$_{2}$TCNQ at 294 K\cite{1DMott1} as shown in Figs.\ref{fig8}(a1) and (b1). 
We also obtain $T$/eV values from the reported lowest excitation energy, 0.694 eV, as $T=0.107$ eV for (i) 
and $T=0.111$ eV for (ii), respectively. Using this, we can check that a theoretical half width of the Lorentzian spectrum 
in Eq.(\ref{eq39}), $2\gamma\sim 0.1$ eV, is consistent with the reported fitting parameter of ET-F$_{2}$TCNQ at 294 K. 
Newly introducing $\gamma$, we subsequently calculate $\Delta\sigma(\omega)$, which denotes the change of 
optical conductivity spectra with and without $\varepsilon$, where $\varepsilon$ is related to the magnitude of 
a modulated electric field (see Eq.(\ref{eq8})). 
Accordingly, $\varepsilon=1.31\times 10^{-2}$ for (i) and $\varepsilon=9.53\times 10^{-3}$ for (ii) are obtained as the best values. 
As clearly seen in Figs.\ref{fig8}(a2) and (b2), although quantitative discrepancies with experimental results still remain 
to some extent, the experimental single {\it plus-minus-plus} structure is completely reproduced. We note that this structure 
is not sufficiently reproduced by finite size calculations within the framework of the exact diagonalization method 
(see Fig.\ref{fig9} in Appendix \ref{AppdxB}). Therefore, above-discussed theoretical spectra with huge system size are essential 
for comparing with experiments. 
\par
To clarify the influences of the many-body effects, we next calculate above same spectra by utilizing a holon-doublon (HD) 
model\cite{HDmodel}, which is a minimal effective model of photoexcitations in a one-dimensional (1D) Mott insulator 
(see Appendix \ref{AppdxA}). This effective model only contains 
the $M=1$ (one HD pair) subspace. As shown in Figs.\ref{fig8}(a1), (a2), (b1), and (b2), the whole spectral structures of the HD model 
almost quantitatively resemble those of previous-mentioned MBWFs results. This similarity is considered to be incidentally occurred 
because of the spectral discrepancies between a MBWFs effective model and HD model at $V_{\alpha\geq 1}=0$ 
in our previous paper (see Fig.8 in Ref.\cite{CHRM}). However, excitonic energy structures of the 
HD model are obviously different from those of the MBWFs results as illustrated in Figs.\ref{fig8}(a3) and (b3). 
Here, the excitonic energy structures are just obtained by calculating eigenenergies of $\widetilde{h}_{\rm ex}^{\pm}$ 
at $N_{\rm ex}=400$ in Eq.(\ref{eq37}). Then, we define the eigenenergies as $\widetilde{E}_{\rm ex}^{\pm}(\nu)$, 
where $1\leq\nu\leq I_{\rm ex}^{\pm}$ and 
$\widetilde{E}_{\rm ex}^{\pm}(1)\leq\cdots\leq\widetilde{E}_{\rm ex}^{\pm}(I_{\rm ex}^{\pm})$.
In the case of the HD model, all the $\nu$th eigenenergies with odd-parity and even-parity are almost completely degenerated. 
Namely, $\widetilde{E}_{\rm ex}^{+}(\nu)\sim\widetilde{E}_{\rm ex}^{-}(\nu)$ at fixed $\nu$ is satisfied for the cases of 
both (i) and (ii). In contrast, for the effective model of MBWFs, such degeneracy is resolved. 
\par
To compare with the previous research\cite{1DMott1}, we focus on the relationship between the lowest two eigenenergies 
$\widetilde{E}_{\rm ex}^{\pm}(\nu=1,2)$ for the effective model of MBWFs. For the case of (i) (see Fig.\ref{fig8}(a3)), 
the odd-parity exciton is clearly bound, while the even-parity exciton is unbound. In addition, the starting energy of 
the even-parity continuum is blue shifted from that of the odd-parity one by $\Delta_{E}$. However, because 
$\Delta_{E}\sim 2$ meV $\sim 0.02T$ is within the evaluated error of our effective models in Sect.\ref{sect2b}, 
we can regard the starting energies between odd-parity and even-parity continuum are subequal. 
Consequently, the excitonic energy structure in the low-energy region is estimated as 
$\widetilde{E}_{\rm ex}^{-}(1)<\widetilde{E}_{\rm ex}^{-}(2)\sim\widetilde{E}_{\rm ex}^{+}(1)\sim\widetilde{E}_{\rm ex}^{+}(2)$, 
which is consistent with the schematic diagram shown in Fig.\ref{fig6}(a). 
On the other hand, in the case of (ii), the inset of Fig.\ref{fig8}(b3) clearly indicates the excitonic energy structure of 
$\widetilde{E}_{\rm ex}^{-}(1)<\widetilde{E}_{\rm ex}^{+}(1)<\widetilde{E}_{\rm ex}^{-}(2)\sim\widetilde{E}_{\rm ex}^{+}(2)$. 
This means that both odd-parity and even-parity excitons are bound and the starting energy of the continuum is almost 
the same for odd-parity and even-parity. 
Extracting a common point from above two relations, 
the even-parity exciton is less tightly bound than the odd-parity exciton due to many-body effects arising 
from the full charge fluctuations. This feature is consistent with the recent experimental fitting parameters of 
ET-F$_{2}$TCNQ at 294 K by using a four-level model\cite{1DMott1} whether or no. 
\par
Accordingly, the many-body effects clearly yields the significant difference of excitonic energy structures between the effective 
model obtained from MBWFs and HD model. Namely, an even-parity excitonic state is less tightly bound than an odd-parity one. 
We discuss the origin of the difference, in what follows, by simplifying the problem with choosing the case of (i). 
Regarding such bound excitons, the most crucial term is the $r(r_{\rm HD})=1$ matrix element of the effective model, namely, 
\begin{equation}
\left(
\widetilde{h}_{\rm ex}^{\pm}
\right)_{1,1} \equiv U-V+\Delta\widetilde{E}^{\pm} \equiv U-\widetilde{V}^{\pm}, 
\label{eq42}
\end{equation}
where $\Delta\widetilde{E}^{\pm}$ describes the correction energy attributed to many-body effects. 
From the previous results shown in Figs.\ref{fig5}(d1) and (e1), 
$\Delta\widetilde{E}^{+}=1.68T>\Delta\widetilde{E}^{-}=1.26T$ can be estimated. This suggests that the renormalized long range 
Coulomb interaction strength satisfies with $-\widetilde{V}^{-}<-\widetilde{V}^{+}$, which roughly means that the attractive 
interaction strength of odd-parity is stronger than that of even-parity at $r(r_{\rm HD})=1$. Within the exact calculations 
of $(\widetilde{h}^{+})_{1,1}-(\widetilde{h}^{-})_{1,1}$ in Eq.(\ref{eq23}) for $N=12, 14, 16$, and 18, we fitted them to a $1/N$-linear 
function and found $(\widetilde{h}^{+})_{1,1}-(\widetilde{h}^{-})_{1,1}\rightarrow 0.291T$ for $N\rightarrow +\infty$. Although the 
$1/N$-dependency of the fitting function should be more carefully discussed to some extent, the value of $0.291T$ roughly denotes 
$\Delta\widetilde{E}^{+}-\Delta\widetilde{E}^{-}=0.291T>0$ for $N\rightarrow +\infty$. This also supports the weakly bound of 
even-parity exciton. 
We note that, for the HD model, $\Delta\widetilde{E}^{\pm}=0$ is always satisfied and, therefore, both an odd- and even-parity 
exciton bound states are completely degenerated. 
%
%
%
\section{Summary}
\label{sect4}
\par
To summarize, we have theoretically investigated excitonic optical conductivity spectra of a Mott insulator with and without 
a modulated electric field at absolute zero temperature. Applying a many-body Wannier functions (MBWFs) method to 
a charge model, which is interpreted as a good effective model of photoexcitations of a one-dimensional (1D) Mott insulator 
in the spin-charge separation picture, we have succeeded in constructing both odd- and even-parity 
one-holon-doublon pair effective models of photoexcitations with including many-body effects at huge system size. 
To validate this, we have studied the photoexcitations of 
ET-F$_{2}$TCNQ, which is a typical 1D Mott insulator. As a result, the theoretical spectra with appropriate 
broadenings qualitatively reproduce the recent experimental data of ET-F$_{2}$TCNQ at 294 K. In addition, obtained 
binding energies of excitons with even-parity are smaller than those with odd-parity due to many-body effects. 
This trend is qualitatively consistent with the analytical results of the experimental spectrum using a simple four-level model, 
which includes two odd-parity and one even-parity excited states, and the ground state. To perform theoretical analyses on 
an experimental spectrum more precisely, an optical conductivity spectrum with a modulated electric field at 4 K is desirable, 
since sharpening of a {\it plus-minus-plus} structure as observed in Figs.\ref{fig8}(a2) and (b2) enables us to obtain 
a more quantitative energy-level structure as well as a more exact parameter set. 
\par
We would like to note that we can further develop our MBWFs method to obtain more highly quantitative estimations comparable 
to experiments. Particularly in this paper, one of the origins of the quantitative mismatch between the theory and experiment 
is considered to be the perturbative treatment of  $H_{V}^{({\rm C})}$ term. 
Although the little discrepancy of excitation energies between exact calculations and 
MBWFs method calculations can be seen (see the filled and empty circles in Figs.\ref{fig6}(b), (c)), the discrepancy possibly 
increases as raising $V$. 
Judging from this, one of our crucial tasks is to establish alternative treatments of including many-body effects 
of $H_{V}^{({\rm C})}$ term beyond our present perturbative approach. 
This also leads to quantitative comparison with the experiments of 1D Mott insulators, such as a Ni-halogen chain compound, 
which is thought to have a large $V$\cite{1DMott3}. 
To produce appropriate MBWFs effective models of photoexcitations for 
more general tight-binding models with higher dimensions, containing two-dimensional Mott insulator, in the thermodynamic limit 
is also one of attracting future works. 
However, our current results are sufficiently useful to understand many-body excitonic excitations in a 1D Mott insulator and 
contribute to reveal the physics of photoinduced phenomena in strongly correlated electron systems. 
\begin{acknowledgments}
This work was supported by JST CREST in Japan (Grant No. JPMJCR1661). 
K.I. was supported by the Grant-in-Aid for Scientific Research from JSPS in Japan (Grant No. JP17K05509). 
T.M. was supported by the Grant-in-Aid for Scientific Research from JSPS in Japan (Grant No. JP20K03801). 
H.O. was supported by the Grant-in-Aid for Scientific Research from JSPS in Japan (Grant No. JP18H01166). 
\end{acknowledgments}
\appendix
\section{Brief introduction of a holon-doublon model and the transformation matrices to construct 
many-body Wannier functions (MBWFs)}
\label{AppdxA}
\par
In this section, using notations of a charge model, we introduce a holon-doublon (HD) model\cite{HDmodel}, which minimally describes 
one-HD pair excited states in a one-dimensional (1D) Mott insulator. Namely, the HD model only contains the $M=1$ subspace 
and neglects all the charge fluctuations arising from $M\geq 2, M=0$ subspaces, where $M$ denotes the number of HD pairs 
in a charge model. 
Then, the Mott-insulator ground state and all the photoexcited states are only included in the subspace of $M=0$ 
and $M=1$, respectively. This approximation is, of course, validated in the limit of $U/T\rightarrow +\infty$. 
Regarding this, in Sects.\ref{sect2} and \ref{sect3}, we compare the results of the effective models constructed by MBWFs with 
those of the HD model to investigate many-body effects originate from full charge fluctuations. 
When one divides the original charge model Hamiltonian $H_{0}^{({\rm C})}+H_{V}^{({\rm C})}$ in Eq.(\ref{eq13}) 
at the $M=1$ subspace into odd-parity part $h_{1}^{-}$ and even-parity part $h_{1}^{+}$, following expressions are obtained. 
\begin{equation}
h_{1}^{-} = 
\left[
\begin{array}{cccccc}
 E_{1}-V_{1}   &      -2T       &             &               &                    &    \bigzerol       \\
  -2T           &  E_{1}-V_{2}  &  \ddots  &               &                    &                       \\
                 &    \ddots      &  \ddots  &  \ddots    &                    &                       \\
                 &                   &  \ddots  &  \ddots    &      -2T         &                       \\
                 &                   &             &    -2T      &  E_{1}-V_{I-1} &       -2T            \\
  \bigzerou  &                   &             &               &      -2T         &     E_{1}-V_{I}
\end{array}
\right], 
\label{eqA1}
\end{equation}
\begin{equation}
h_{1}^{+} = 
\left[
\begin{array}{cccccc}
 E_{1}-V_{1}   &      -2T       &             &               &                    &    \bigzerol       \\
  -2T           &  E_{1}-V_{2}  &  \ddots  &               &                    &                       \\
                 &    \ddots      &  \ddots  &  \ddots    &                    &                       \\
                &                   &  \ddots  &  \ddots    &      -2T         &                       \\
                 &                   &             &    -2T      &  E_{1}-V_{I-1} &    -2\sqrt{2}T     \\
  \bigzerou  &                   &             &               &   -2\sqrt{2}T  &     E_{1}-V_{I}
\end{array}
\right]. 
\label{eqA2} 
\end{equation}
These are derived by calculating all the matrix elements of $\left(h_{1}^{p}\right)_{r^{\prime},r}=
\langle r_{1}^{\prime p}(0)|\left(
H_{0}^{({\rm C})}+H_{V}^{({\rm C})}
\right)|r_{1}^{p}(0)\rangle$ ($p=\pm 1$) where $|r_{1}^{p}(0)\rangle$ is in Eq.(\ref{eq14}) and the subscripts $r^{\prime},r$ naturally 
have one-to-one correspondence with a relative distance between a H and D, $r_{\rm HD}$. Then, the dimension of the matrix, 
$h_{1}^{-}$ ($h_{1}^{+}$), is $I\equiv N/2-1$ ($N/2$) and $E_{1}\equiv U+N\sum_{\alpha}V_{\alpha}$. For the ground state, 
all the sites are singly occupied by electrons and, therefore, its energy is $E_{0}\equiv N\sum_{\alpha}V_{\alpha}$. 
Analytically calculating the inverse matrices of the unitary transformations of diagonalizing $h_{1}^{\pm}$ for vanishing 
all the $V_{\alpha}$, we decide the unitary transformations of constructing MBWFs in this paper as follows. 
\begin{align}
V_{r\mu}^{{\rm tr}+} &= 
\begin{cases}
  \displaystyle{\sqrt{\frac{2}{N}}\sin\left[\frac{\pi}{N}r(2\mu-1)\right]}  & (r=N/2) \\
  \displaystyle{\frac{2}{\sqrt{N}}\sin\left[\frac{\pi}{N}r(2\mu-1)\right]}  & (1\leq r\leq N/2-1)
\end{cases}, 
\label{eqA3} \\
V_{r\mu}^{{\rm tr}-} &= \frac{2}{\sqrt{N}}
\sin\left[\frac{2\pi}{N}r\mu\right] \quad (1\leq r\leq N/2-1).
\label{eqA4}
\end{align}
Applying the definitions in Eqs.(\ref{eq17}) and (\ref{eq19}) to $h_{1}^{\pm}$, 
$|\widetilde{\Phi}_{r}^{\pm}(0)\rangle=|r_{1}^{\pm}(0)\rangle$ is a trivial resultant. From Eq.(\ref{eq20}), 
$J_{rg}^{-}=2c_{\rm S}(1)\delta_{r,1}$ is also derived. Replacing $\widetilde{h}_{\rm ex}^{\pm}$ with 
$h_{1}^{\pm}-E_{0}$ at $N=N_{\rm ex}$ in Eqs.(\ref{eq37}), (\ref{eq38}), (\ref{eq39}), and (\ref{eq41}), the optical spectra with and 
without a modulated electric field for the HD model are calculated with $|J_{\rm ex}\rangle$ in Eq.(\ref{eq40}) of 
\begin{align}
(J_{\rm ex})_{rg}^{+} &= 0 \quad (1\leq r\leq I_{\rm ex}^{+}=N_{\rm ex}/2), 
\label{eqA5} \\
(J_{\rm ex})_{rg}^{-} &\equiv 
\begin{cases}
  2c_{\rm S}(1)  & (r=1) \\
         0           & (2\leq r\leq I_{\rm ex}^{-}=N_{\rm ex}/2-1)
\end{cases}. 
\label{eqA6}
\end{align}
\section{Exact diagonalization analysis for optical conductivity spectra with and without a modulated electric field}
\label{AppdxB}
%
%
\par
\begin{figure}[t]
\begin{center}
\includegraphics[width=8cm,keepaspectratio]{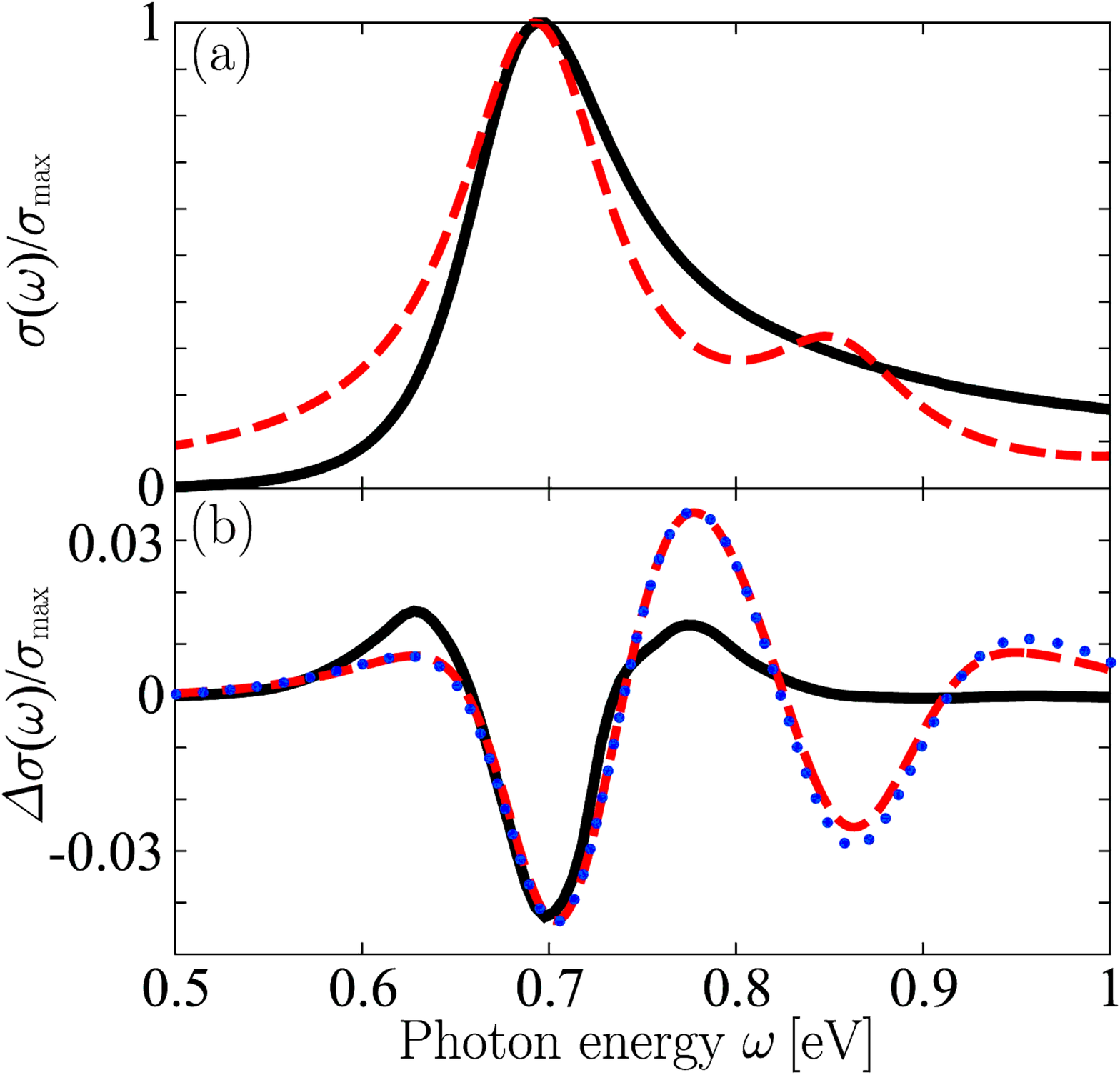}
\caption{Optical spectra of $N=16, U=10T, V_{1}=V=2.8T, V_{\alpha\geq 2}=0$ in Eqs.(\ref{eqB1}) and (\ref{eqB2}). 
The solid lines and the others are 
experimental data of ET-F$_{2}$TCNQ at 294 K\cite{1DMott1} and 
theoretical results of a charge model with $M_{\rm max}=N/2=8$ at absolute zero temperature, respectively. 
The theoretical spectrum, which limits $H_{\phi}^{({\rm C})}$ in the $M=1$ subspace, is demonstrated as the dotted line. 
From the theoretical lowest excitation energy corresponding to the first peak of $\sigma(\omega)$, $T=0.110$ eV is estimated. 
Spectral broadenings are $\gamma=0.45T$ for the dashed line in (a), $\gamma=0.63T$ for the dashed line in (b), and 
$\gamma=0.665T$ for the dotted line in (b). 
In (b), 
$\varepsilon=2.0\times 10^{-2}$ ($E_{\rm amp}=238$ kV/cm) for the dashed line 
and 
$\varepsilon=4.0\times 10^{-2}$ ($E_{\rm amp}=477$ kV/cm) for the dotted line. 
}
\label{fig9}
\end{center}
\end{figure}
\par
\par
In this section, exemplifying the case of (i) in Sect.\ref{sect3} ($V_{1}=V=2.8T$ and $V_{\alpha\geq 2}=0$), 
we perform finite-size exact diagonalization calculations of optical conductivity spectra of a charge model with and without 
$H_{\phi}^{({\rm C})}$ term (see Eqs.(\ref{eq4}) and (\ref{eq10})), which generates a modulated electric field. 
Using $H^{({\rm C})}$ in Eq.(\ref{eq13}) and $J^{({\rm C})}$ in Eq.(\ref{eq11}), 
we rewrite some formulations needed to calculate the optical spectra in this section. Namely, for given photon energy $\omega>0$, 
\begin{align}
\sigma(\omega) &= -\frac{1}{N\omega}{\rm Im}\left[
\langle g|J^{({\rm C})}\frac{1}{\omega+i\gamma+E_{g}-H^{({\rm C})}}J^{({\rm C})}|g\rangle
\right], 
\label{eqB1} \\
\Delta\sigma(\omega) &\equiv \sigma(\omega)|_{\varepsilon\neq 0} - \sigma(\omega)|_{\varepsilon=0}, 
\label{eqB2}
\end{align}
where $H^{({\rm C})}|_{\varepsilon=0}|g\rangle=
\left(
H_{0}^{({\rm C})}+H_{V}^{({\rm C})}
\right)|g\rangle\equiv E_{g}|g\rangle$, $\gamma$ represents an artificial broadening, and $\varepsilon$ is in Eq.(\ref{eq8}). 
In order to coinstantaneously compute $\sigma(\omega)|_{\varepsilon\neq 0}$ and $\sigma(\omega)|_{\varepsilon=0}$, we treat 
a bare basis representation in the absence of both a parity inversion and translational symmetry. Because of this, the 
calculable maximum size is limited to $N=16$ for computational problem and all the spectra are computed by 
a continued fraction method\cite{CFRA}. The results, comparing with the experiments of ET-F$_{2}$TCNQ at 294 K\cite{1DMott1}, 
are illustrated in Fig.\ref{fig9}. 
The values of $\gamma$,  $T$/eV, and $\varepsilon$ are determined in the same manner as mentioned in Sect.\ref{sect3}. 
As seen in Fig.\ref{fig9}, due to finite size effects, few separated spectral peaks and dips still remain in spite of 
sufficiently large broadenings. Particularly, theoretical $\Delta\sigma(\omega)$ in the $\omega\gtrsim 0.8$ eV region, 
the second dip and third peak clearly appear, compared with the corresponding MBWFs results in Fig.\ref{fig8}(a2). 
Namely, in that region, the spectral magnitudes for $N=16$ are large and far from the experimental spectrum. In addition, 
although the exact excitonic energy structure at $N=16$ also contains non-negligible finite size effects (see Fig.\ref{fig6}), 
the second peak position of theoretical $\Delta\sigma(\omega)$ accidentally coincides with that of experiment. 
Regarding this, the experimental spectra are only comparable to corresponding theoretical spectra of adequately large system size 
with negligible finite size effects as shown in Fig.\ref{fig8}, for instance. 


\begin{thebibliography}{99}
\bibitem{Comment0} As mentioned in our previous paper\cite{CHRM}, this theoretical approach is related to a configuration 
interaction (CI) method\cite{CI} or Tamm-Dancoff approximation\cite{TDA} in the literature. The CI method has been widely used 
as one of the typical treatments to perturbatively calculate highly excited many-electron basis states\cite{CIEX1,CIEX2}.
\end{thebibliography}
\end{document}